\documentclass[aps, prd, reprint,10pt, notitlepage, a4paper,floats, amsmath, amssymb, amsfonts,superscriptaddress,showpacs, showkeys,nofootinbib,longbibliography]{revtex4-2}

\usepackage[varg]{txfonts}
\usepackage[colorlinks=true,linkcolor=blue,linktocpage=true,anchorcolor=black,citecolor=blue,filecolor=black,menucolor=black, urlcolor=blue]{hyperref}
\usepackage{graphicx,psfrag}
\usepackage{grffile} % Package grffile changes the algorithm to check for known file extensions, like .jpg
\usepackage{mathrsfs}
\usepackage{amsmath,amsfonts,amssymb}
\usepackage{tasks} % for horizontal lists
\usepackage{multirow}
\usepackage{comment}
\usepackage{bm}
\usepackage{paralist}
\usepackage{ulem}
\usepackage{booktabs}
\usepackage{xcolor,color}
\usepackage{caption}
\usepackage{subcaption}
\usepackage{multirow} % for multirow in tables
\usepackage{enumitem}
\usepackage{bm} %% package for bold fonting greek letters in math mode
\usepackage[T1]{fontenc}
\usepackage{soul}
\usepackage[mathlines]{lineno}% Enable numbering of text and display math

\def\Msun{M_\odot}

\def\XPHM{\textsc{XPHM}}
\def\XOFourA{\textsc{XO4a}}
\def\TPHM{\textsc{TPHM}}
\def\SEOB{{\textsc{SEOB}}}
\def\NRsur{{\textsc{NRSur7dq4}}}
\def\SXS{\texttt{SXS}}
\newcommand{\BAM}[1]{{\texttt{CF\_}}{\texttt{#1}}}
\newcommand{\sxsSim}[1]{{\texttt{SXS:\,}}{\texttt{#1}}}

\newcommand{\be}{\begin{equation}}
\newcommand{\ee}{\end{equation}}
\newcommand{\ba}{\begin{align}}
\newcommand{\ea}{\end{align}}
\newcommand{\model}{\mathfrak{M}}
\newcommand{\mismatch}{\mathcal{M}}

\newcommand{\ord}{\mathcal{O}}
\newcommand{\f}{\frac}

\usepackage{color}
\definecolor{cyan}{rgb}{0,0.9,0.9}
\definecolor{orange}{rgb}{0.9,0.5,0}
\definecolor{magenta}{rgb}{1,0,1}
\definecolor{purple}{rgb}{0.8,0.4,0.8}
\definecolor{gray}{rgb}{0.8242,0.8242,0.8242}
\definecolor{dodgerblue}{rgb}{0.12, 0.56, 1.0}

\newcommand{\new}[1]{\textcolor{black}{#1}} %teal
\newcommand{\chip}{\chi_\text{p}}

\newcommand{\chiperp}{\chi_{\perp}}
\newcommand{\chieff}{\chi_{\mathrm{eff}}}

\newcommand{\modelname}[1]{\textsc{#1}}
\newcommand{\Mtot}{M}
\newcommand{\insp}{\text{I}}
\newcommand{\rat}{\mathfrak{r}}
\newcommand{\wid}{\mathfrak{w}}

\begin{document}

\title{Waging a Campaign: Results from an Injection/Recovery Study involving 35 numerical Relativity Simulations and three Waveform Models}

\author{Sarp \surname{Ak\c{c}ay}}
\affiliation{University College Dublin, Belfield, D4, Dublin, Ireland}
\author{Charlie \surname{Hoy}}
\affiliation{Institute of Cosmology and Gravitation, University of Portsmouth, Portsmouth, PO1 3FX, UK}
\author{Jake \surname{Mac Uilliam}}
\affiliation{University College Dublin, Belfield, D4, Dublin, Ireland}
\affiliation{Irish Centre for High-End Computing, 2, 7/F, The Tower, Trinity Technology and Enterprise Campus, Grand Canal Dock, D2, Dublin, Ireland}

\date{\today}% It is always \today, today,

\begin{abstract}
We present Bayesian inference results from an extensive injection-recovery campaign to test the validity of three state of the art quasicircular gravitational waveform models: \modelname{SEOBNRv5PHM}, \modelname{IMRPhenomTPHM}, \modelname{IMRPhenomXPHM}, the latter with the \modelname{SpinTaylorT4} implementation for its precession dynamics. We analyze
35 strongly precessing binary black hole numerical relativity simulations with all available harmonic content.
Ten simulations have a mass ratio of $4:1$ and five, mass ratio of $8:1$.
Overall, we find that \modelname{SEOBNRv5PHM} is the most consistent model to numerical relativity, with the majority of true source properties lying within the inferred 90\% credible interval. However, we find that none of the models can reliably infer the true source properties for binaries with mass ratio $8:1$ systems. We additionally conduct inspiral-merger-ringdown (IMR) consistency tests to determine if our chosen state of the art waveform models infer consistent properties when analysing only the inspiral (low frequency) and ringdown (high frequency) portions of the signal.
For the simulations considered in this work, we find that the IMR consistency test depends on the frequency that separates the inspiral and ringdown regimes. For two sensible choices of the cutoff frequency, we report that \modelname{IMRPhenomXPHM} can produce false GR deviations. Meanwhile, we find that \modelname{IMRPhenomTPHM} is the most reliable model under the IMR consistency test.
Finally, we re-analyze the same 35 simulations, but this time we incorporate model accuracy into our Bayesian inference. Consistent with the work in Hoy et al. 2024 [arXiv: 2409.19404 [gr-qc]], we find that this approach generally
yields more accurate inferred properties for binary black holes with less biases compared 
to methods that combine model-dependent posterior distributions based on their evidence, or with equal weight.

\end{abstract}

\maketitle

\section{Introduction}
With a decade of observations, $\ord(100)$ confirmed gravitational-wave (GW) detections \cite{LIGOScientific:2021djp, Nitz:2021zwj,Olsen:2022pin,Mehta:2023zlk,Wadekar:2023gea} 
and $\ord(200)$ potential GW candidates \cite{gracedb} under ``its belt'',
GW astronomy has matured into an established branch of astronomy.
At the forefront of this endeavour are the GW interferometers~\cite{TheLIGOScientific:2014jea,acernese2014advanced,KAGRA:2020tym} (IFOs) operated by
the LIGO-Virgo-KAGRA collaboration (LVK), with plans for an additional detector in India~\cite{dcc:M1100296}.
An overwhelming majority of the detected events have been sourced
by the inspiral, merger and ringdown (IMR) of binary black hole systems (BBHs) \cite{LIGOScientific:2018mvr, Abbott:2020niy, LIGOScientific:2021usb, LIGOScientific:2021djp}.
 Assuming vacuum general relativity and negligible orbital eccentricity at the time the GW
 signal enters the sensitive region of the IFOs, ca. $20\,$Hz, each system should be fully characterized by 15 parameters: eight intrinsic (two masses and two Euclidean spin vectors) and seven extrinsic (sky position, orbit orientation, time and phase shifts).

Gravitational-wave Bayesian analyses are crucial for extracting source properties from noisy data. 
Typical Bayesian analyses rely on stochastic sampling of the
15-dimensional parameter space~\cite{Veitch:2014wba} with $\sim O(10^{7})$ waveform evaluations per analysis. This implies that fast and faithful GW models are needed.
The faithfulness of a GW model is gauged with respect to state of the art numerical relativity (NR)
simulations, such as those produced by the SXS \cite{Mroue:2013xna, Boyle:2019kee, SXS:catalog, Scheel:2025jct}, BAM \cite{Hamilton:2023qkv, hamilton_eleanor_2023_7673796, hamilton_eleanor_2023_7677297}, RIT \cite{Healy:2017psd, Healy:2019jyf, Healy:2020vre, Healy:2022wdn}, NINJA \cite{Ajith:2012az}, NRAR \cite{Hinder:2013oqa}, MAYA collaborations \cite{Jani:2016wkt, Ferguson:2023vta} as well as 
the GR-Athena++ catalog \cite{Rashti:2024yoc} and the works of Refs~\cite{Huerta:2019oxn}.
Each NR simulation is expensive to produce, taking days to months to
complete on a high performance computing clusters. This means that direct use of NR waveforms is
not feasible for GW data analysis (although see e.g. Ref.~\cite{Healy:2020jjs}).
Instead, waveform models based on analytical, semi-analytical and
phenomenological methods are employed in parameter estimation analyses.
Such methods allow waveform generation to be fast at the expense of faithfulness
to NR simulations, which may cause biases in parameter estimation.

It is customary to perform injection and recovery Bayesian analyses to assess when a waveform model may produce biased estimates for the true underlying source properties
(see, e.g., Refs. \cite{Pratten:2020ceb, Estelles:2021gvs, Gamba:2021ydi, Ramos-Buades:2023ehm}).
These are procedures whereby a template waveform, usually from an NR simulation,
is injected into a single or multiple IFOs and the resulting signal analyzed with
different waveform models through Bayesian methods~\cite[see e.g.][]{TheLIGOScientific:2016wfe}.
In order to solely focus on systematic biases induced by waveform models, the template waveforms are generally injected into zero detector noise,
though the dependence of the likelihood on detector noise is retained (see Sec.~\ref{Sec:injection_primer}).
Such studies usually involve a few injections as they are computationally expensive.

Here, we undertake a systematic survey of waveform model recovery for numerical relativity injections,
prioritizing strongly precessing BBHs. 
A similar injection campaign was conducted in Ref.~\cite{Abbott:2016wiq} 
to assess the systematic biases in the waveform models used in GW150914.
That particular study focused mostly on the injection-recovery of non-precessing systems.
Our singling out strongly precessing cases for our campaign is motivated by:
\begin{enumerate}[label=(\roman*)]
 \item 
Waveform systematics for non-precessing, quasicircular binary black holes are well ``under control''
with the state of the art models achieving $10^{-3}$ level of faithfulness (median) to numerical relativity simulations \cite{Cotesta:2018fcv, Varma:2018mmi, Estelles:2020twz, Pratten:2020fqn, Garcia-Quiros:2020qpx, Riemenschneider:2021ppj, Pompili:2023tna, Nagar:2023zxh}. 
Models older than these were found to cause no systematic biases 
in the recovery of the systems with injected parameters similar to GW150914 \cite{Abbott:2016wiq},
i.e., either no precession or weak precession \cite{TheLIGOScientific:2016wfe}.
\item 
Precessing waveform models now attain $\lesssim 10^{-2.5}$ faithfulness (median)
for systems with weak to mild precession 
(see, e.g., Ref.~\cite{MacUilliam:2024oif}).
\item
Last, although they are thought to be rare~\cite{LIGOScientific:2021psn,Hoy:2024qpy}, (vacuum) binary black hole systems exhibiting strong spin precession are astrophysically interesting as they may point to a different formation channel~\cite{Kalogera:1999tq, Mandel:2009nx, Gerosa:2018wbw,Rodriguez:2016vmx,Belczynski:2014iua,Mandel:2018hfr}. Therefore, the understanding of the expected level of bias in the inferred source properties for these systems will have implications on our ability to infer the binary black hole history.
\end{enumerate}
So far, GW200129\_065458 has been the only observation which exhibits spin-precession
\new{under the assumption of circular orbits} \cite{LIGOScientific:2021djp, Hannam:2022pit, Macas:2023wiw, Estelles:2025zah}.
The rarity of detecting GW200129\_065458-like systems has been quantified by Ref.~\cite{Hoy:2024qpy}
to be once out of every $\lesssim 50$ events.
At the time of GW200129\_065458, it was shown in Ref.~\cite{Hannam:2022pit} that the numerical relativity surrogate model, \textsc{NRSur7dq4} \cite{Varma:2019csw}, was the least likely to produce biased parameter
estimates for a strongly precessing BBH with dimensionless primary spin magnitude of $0.9$ and spin tilt angle of nearly $90^\circ$ (possibly due to incorporating multipole-asymmetries~\cite{Kolitsidou:2024vub} \new{also confirmed recently by Ref.~\cite{Estelles:2025zah}}). \new{N.B.: the consensus on the detectability of spin precession for this event is not universal and can change depending on the assumptions made \cite{Payne:2022spz,Gupte:2024jfe,Fernando:2024zcd, Planas:2025jny}}. %which disagree with some of these findings.

Therefore, our goal is to assess the performance of the latest precessing waveform models developed since 2020, specifically, \modelname{SEOBNRv5PHM}~\cite{Ramos-Buades:2023ehm} (hereafter \SEOB),
\modelname{IMRPhenomTPHM} (thenceforth \TPHM)~\cite{Estelles:2021gvs} and \modelname{IMRPhenomXPHM}~\cite{Pratten:2020ceb,Colleoni:2024knd} (henceforward \XPHM), for Bayesian analyses.
Given that the primary investigation of the only published O4 event, GW230529\_181500, uses \modelname{SEOBNRv5PHM}
and \modelname{IMRPhenomXPHM} \cite{LIGOScientific:2024elc}, our choosing of these
two particular waveform models is appropriate (also motivated by Ref.~\cite{Dhani:2024jja}).
We also include \modelname{IMRPhenomTPHM} in our work as it is represents a relatively fast time-domain model, which bridge the gap between the time-domain effective one-body~\cite{Bohe:2016gbl,Babak:2016tgq}, and the frequency-domain phenomenological approaches~\cite{Husa:2015iqa,Khan:2015jqa} both in terms of accuracy \cite{MacUilliam:2024oif} and speed.
Finally, we also employ \modelname{IMRPhenomXO4a} \cite{Thompson:2023ase} for some selected cases for which \XPHM{} exhibits poor recovery performance.
Note that we work with the version of \XPHM{} with the Spin-Taylor prescription for the spin dynamics  \cite{Colleoni:2024knd}, which is more accurate than its MSA version \cite{Chatziioannou:2017tdw,Pratten:2020ceb}.

Our assessment involves the injection of 35 numerical relativity waveforms from the 
\SXS\footnote{The updated and enlarged 2025 \SXS{} catalog \cite{Scheel:2025jct}
became available as we were completing this work. Therefore, we employ waveforms from the 2019 catalog.}-\texttt{BAM} databases \cite{Boyle:2019kee, Hamilton:2023qkv}
into zero-noise {two-detector} advanced LIGO network with design {O4} amplitude spectral densities~\cite{O4PSD} 
(ASD).
The simulations are divided into three subsets of size 10 and a fourth subset of size five.
The 10-element subsets are separated by their mass ratios: $1:1, 2:1, 4:1$ with the five-element
subset containing simulations with mass ratio $8:1$.
We have chosen simulations that exhibit strong precession, with extrinsic
parameters chosen such that a quarter of the accumulated network signal to noise ratio (SNR)
is purely due to the precessing part of the waveform \cite{Fairhurst:2019vut}.
This ensures that the precession signal is not diminished due to the orbital configuration of the BBHs with respect
to the detector network.

Another important assessment of waveform systematics is the inspiral-merger-ringdown (IMR) consistency tests
\cite{Ghosh:2016qgn, Ghosh:2017gfp, LIGOScientific:2019fpa},
whereby the final mass and spin of the merger-product black hole is computed from both the inspiral
(pre-merger) and the ringdown (post-merger) phases independently. {Assuming general relativity to be the correct theory of gravity}, 
a robust waveform model should produce
consistent results for the inferred values of the final mass and the spin.
This test is routinely performed on GW events \cite{LIGOScientific:2019fpa, LIGOScientific:2020tif, LIGOScientific:2021sio}.
Here, we perform a systematic investigation of the waveform model robustness under the IMR consistency tests
over our sample of 35 numerical relativity signals. 
As the IMR consistency tests are computationally expensive, we employ the fastest of the three waveform models considered, \XPHM, for the entire sample set.
We re-perform the IMR tests with \modelname{SEOB}, \modelname{TPHM} and \modelname{IMRPhenomXO4a} for cases in which \XPHM{} shows
noteworthy biases.
We then repeat the IMR consistency test usinga different frequency to distinguish between the inspiral and ringdown phases to assess how this change affects our results. We again employ \modelname{XPHM} for all 35 cases,
and perform supplementary runs with the other waveform models as needed.
As far as we are aware, our undertaking here is the most systematic and comprehensive
application of the IMR consistency test.

This work provides us with model-dependent posterior distributions that can be
combined into multi-model posteriors to marginalize over waveform uncertainty.
The standard way of doing is to assign equal weights to each model's 
posterior as has been done routinely by the LVK \cite{LIGOScientific:2021djp}.
Alternatively, Ref.~\cite{Ashton:2019leq} suggested combining the posteriors
based on the Bayesian evidence obtained by each waveform model. We will refer to this
as the evidence-informed method.
In Ref.~\cite{Hoy:2024vpc}, we proposed yet another method for combining multi-model posteriors
based on each waveform model's
faithfulness to NR in the parameter space explored by the stochastic sampler(s). The large number of posteriors generated by our study
allows us to make a thorough comparison between these three methods for
combining posteriors.
As the method in Ref.~\cite{Hoy:2024vpc}, dubbed NR-informed, is currently restricted to BBHs between mass ratios of $1:1$ to $4:1$, we do not apply it to the five \texttt{BAM}
injections of mass ratio $8:1$.
To our knowledge, our work here is the \emph{first} large-scale systematic study of the latter two methods for combining posteriors from multiple precessing waveform models.

\begin{figure*}
 \includegraphics[width=0.49\textwidth]{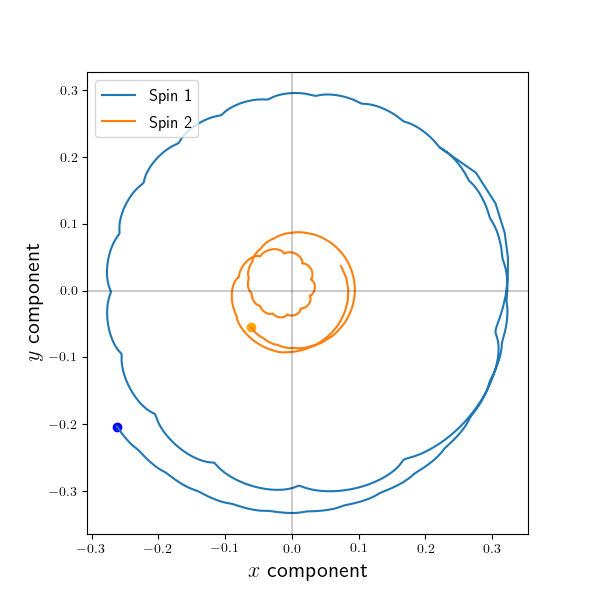}
 \includegraphics[width=0.49\textwidth]{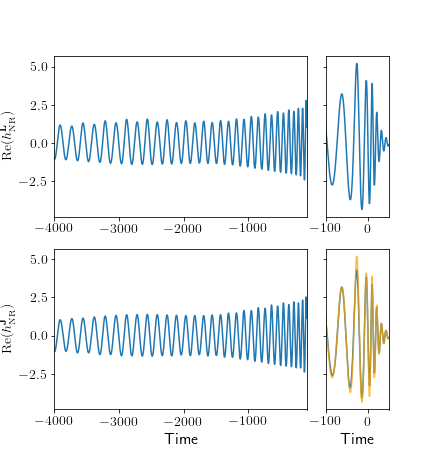}
\caption{Left panel: The precession of $\mathbf{S}_1$ and $\mathbf{S}_2$ around the total angular momentum 
vector $\mathbf{J}$ which points out of the page through the origin.
As the spin vectors precess, they trace out precession cones around $\mathbf{J}$ whose projections
are plotted in the figure as the blue and orange curves.
The small fluctuations in the trajectories are due to nutation. 
As the mass ratio of this system (\texttt{SXS:BBH:1200}) is $2:1$ and the dimensionless spins are equal, the magnitude of $\mathbf{S}_1$
is approximately $2^2=4$ times larger than the magnitude of $\mathbf{S}_2$. The dots mark the initial projections of the spin vectors.
Right panel: The numerical relativity waveform strain $\mathrm{Re}(h_\text{NR}(t))$ for this simulation seen both in the 
$\mathbf{L}$ frame (top row) and the $\mathbf{J}$ frame (bottom row). In this figure, the time units have been geometrized (adimensionalized) via $G=c=M=1$. For $M=53.6\Msun$ given for this system in
Table~\ref{tab:SXS_sim_params}, a time interval of $4000M$ converts to $1.06\,$ second. Though the waveforms in the
two frames look very similar, they are not identical as can be seen in the lower right subfigure where
we overlaid the orange $\mathbf{L}$-frame merger-ringdown waveform on top of its $\mathbf{J}$-frame counterpart (blue).}
\label{fig:Precessing_spins}
\end{figure*}

This article is organized as follows: in Sec.~\ref{Sec:injection_primer}, we present a general introduction to injection-recovery
studies and an overview of the specific choices and settings for the runs that we performed.
Sec.~\ref{Sec:recovery_3models} showcases the results of the recoveries by 
\modelname{SEOBNRv5PHM}, \modelname{IMRPhenomTPHM} and \modelname{IMRPhenomXPHM}
for the 35 numerical relativity injections.
Sec.~\ref{Sec:IMR_tests} reports on the results of our IMR consistency tests, performed on \modelname{IMRPhenomXPHM} for all 35 simulations with additional results presented for all three models for eight cases.
Sec.~\ref{Sec:NR_informed_PE} compares the results of our NR-informed parameter estimation runs with those
obtained via the evidence-informed and standard methods.
We summarize our most important findings in Sec.~\ref{Sec:discussion} and discuss the implications.
All the relevant parameter, data and configuration files have been collected in
a \texttt{git} repository which can be accessed at \url{https://github.com/akcays2/Injection_Campaign} \cite{campaign}.

\section{Physical set up and Notation}\label{Sec:notation}

A BBH system comprised of two compact objects with masses $m_{1}$ and $m_{2}$ has total mass $\Mtot=m_1+m_2$, chirp mass $M_c = (m_{1}m_{2})^{3/5} / \Mtot^{1/5}$ and symmetric mass ratio $\eta = m_{1}m_{2} / \Mtot^{2}$. The ratio of binary masses can either be $\ge 1$, defined by the large mass ratio $Q:= m_1/m_2$, or $\le 1$, defined by the small mass ratio $q=1/Q$.
We quote all masses and frequencies in the detector frame. Source-frame masses are related to their detector frame quantities via $\text{mass}_s = (1+z)^{-1} \text{(mass)}$,
where $z$ is the cosmological redshift of the source.
We use geometrized units $G=c=1$ except when quoting values for the total binary mass which we do in units of solar masses and the GW frequency $f$ which will be given in Hertz.

The orbital angular momentum of the binary is given by $\mathbf{L}=L\, \hat{\mathbf{L}}=\mathbf{r}\times\mathbf{p}$ where $\mathbf{r},\mathbf{p}$ are the relative separation
and momentum of the binary respectively, and $L=\vert\mathbf{L}\vert$.
The black hole spins are represented by the Euclidean three-vectors $\mathbf{S}_i$ which have magnitudes
$\vert\mathbf{S}_i\vert=m_i^2 a_i$ with $0\le a_i < 1$ for $i=1,2$,
where we deliberately avoid the extremal Kerr spin limit $a_{i}=1$.
$\theta_{1}, \theta_2$ denote the spin tilt angles with respect to a reference axis which is
taken to be the direction of the \new{Newtonian} orbital angular momentum vector at some reference
frequency, e.g., $20\,$Hz \new{\cite{Schmidt:2017btt}}. Similarly, $\phi_{1},\phi_2$ represent the azimuthal spin angles in the
orbital plane with $\Delta\phi:\phi_2-\phi_1$.

The phenomenology of spin precession in compact binary inspirals was thoroughly
exposed in the seminal work of Ref.~\cite{Apostolatos:1994mx}. 
Using the large disparity between the orbital, precession and radiation reaction timescales, 
and the assumption of conserved total angular momentum, $\mathbf{J}=\mathbf{L}+\mathbf{S}_1+\mathbf{S}_2$, a picture emerges
whereby $\mathbf{L},\mathbf{S}_1,\mathbf{S}_2$ all precess
around $\mathbf{J}$, tracing out cones over a precession period.
We show the two-dimensional projection of two such cones for the case of the $Q=2$
binary black hole simulation (\texttt{SXS:BBH:1200})~\cite{Boyle:2019kee} in the left panel of Fig.~\ref{fig:Precessing_spins}
(see also e.g., Fig.~1 of Ref.~\cite{Ossokine:2015vda}).
The relevant parameters of this system can be found in Tables~\ref{tab:SXS_sim_params} and \ref{tab:SXS_extrinsic_params}.
In the right panel of the figure, we show the GW signal produced by this merging system as
seen in the $\mathbf{L}$ frame and in the
$\mathbf{J}$ frame. The $\mathbf{L}$ frame is standard for NR, and in the $\mathbf{J}$ frame the amplitude modulations due to precession are in general less pronounced. However, this is hard to discern for this particular case because the precession cone opening angles
for $\mathbf{L}$ ($\mathbf{J}$) are small with respect to their frames $\mathbf{L}_0$ ($\mathbf{J}_0$).
The differences between the two waveforms become more pronounced when the spins
point mostly in the opposite direction to $\mathbf{L}_0$ (see, e.g., Fig.~3 of Ref.~\cite{Akcay:2020qrj}.)

Often it is convenient to describe the individual spin vectors by effective inspiral and precessing spins~\cite{Damour:2001tu, Racine:2008qv, Hannam:2013oca, Purrer:2013ojf,Hannam:2013oca, Schmidt:2014iyl}. The effective inspiral spin is defined as
\be
\chi_\text{eff} = \f{1}{1+q}(a_1 \cos\theta_1+ q a_2\cos\theta_2) ,
\label{eq:chi_eff}
\ee
and the effective precession spin
\begin{equation}
  \chip=\max\left(a_1\sin\theta_1,q\,\frac{4q+3}{4+3q}\,a_2\sin\theta_2\right). \label{eq:chi_p}
\end{equation}
These obey the bounds $-1< \chieff <1$ and $0\le \chip <1$.
Additionally, we can define the effective parallel spin \cite{Hoy:2024vpc}
\be
\chi_\parallel := \f{1}{(1+q)^2}(a_1 \cos\theta_1+ q^2 a_2\cos\theta_2) 
\label{eq:chi_par}
\ee
and the effective perpendicular spin \cite{Akcay:2020qrj,MacUilliam:2024oif,Hoy:2024vpc}
\begin{equation}
  \chiperp=\frac{\vert \mathbf{S}_{1,\perp}+\mathbf{S}_{2,\perp} \vert}{M^2},\label{eq:chiperp}
\end{equation}
where $\mathbf{S}_{i,\perp}=a_i m_i^2(\sin\theta_i\cos\phi_i, \sin\theta_i\sin\phi_i ,0)^T$ for $i=1,2$.

The masses and spins combine to make up the eight intrinsic parameters, $\boldsymbol{\lambda}_\text{int}$, of a given BBH.
There are additionally seven extrinsic parameters in the case of quasicircular (quasispherical) inspirals:
the luminosity distance $d_\text{L}$, the right ascension and declination angles $\{\alpha,\delta\}$,
\new{an overall constant time shift $t_\text{ref}$; and
polar (inclination), azimuthal and polarization angles $\{\iota,\varphi_\text{ref},\psi\}$ parametrizing the orientation of the orbit and the polarization of the GWs with respect to the detector frame}.

\new{
With the reelvant parameters of the system specified, the gravitational-wave strain
in the time domain can be written in its multipolar form as
\be
h(t)=\f{1}{d_\text{L}} \sum_{\ell,m} h_{\ell m}(t, \boldsymbol{\lambda}_\text{int}){}_{-2}Y^{\ell m}(\iota,\varphi_\text{ref}),
\label{eq:h_lm}
\ee
where ${}_{-2}Y^{\ell m}$ are spin-weighted spherical harmonics and $h_{\ell m}$ are the
GW multipoles. The three models of interest here use similar prescriptions
to frame-rotate the co-precessing multipoles $h^\text{coprec}_{\ell m}$ to $h_{\ell m}$
using the orientation of $\mathbf{L}(t)$ with respect to $\mathbf{J}_0$ during the inspiral
to determine these Euler angles \cite{Schmidt:2010it, Schmidt:2012rh, Boyle:2011gg}. The models
use more differing approximations to extend the Euler angles into the merger-ringdown
regime.
The co-precessing multipoles can be modelled using the aligned-spin (non-precessing) multipoles as is done, e.g., for \TPHM{} and \XPHM.
\SEOB{} has a more sophisticated prescription whereby partial precessing-spin information
is incorporated into the EOB Hamiltonian used in the co-precessing frame \cite{Khalil:2023kep}.
An important commonality in all three models is that they all employ the so-called
multipole symmetry for their co-precessing multipoles, i.e.,
\be
h^\text{coprec}_{\ell, -m} = (-1)^\ell \left[h^\text{coprec}_{\ell m}\right]^\ast
\label{eq:h_coprec_symmetry},
\ee
where $\ast$ denotes complex conjugation.
}

\section{A Primer on Injection/Recovery Studies}
\label{Sec:injection_primer}
The accuracy of a GW model is commonly assessed by performing Bayesian inference on a 
simulated GW signal, $h$, produced from a coalescing binary of known parameters
$\boldsymbol{\lambda}_{\mathrm{inj}}$~\cite[see e.g.][]{Varma:2019csw,Ramos-Buades:2023ehm,Thompson:2023ase,Colleoni:2024knd};
Bayesian inference is the process of estimating a model-dependent
{\textit{posterior probability distribution}}, which describes the probability of the binary having
a specific set of properties $\boldsymbol{\lambda} = \{\lambda_{1}, \lambda_{2}, ..., \lambda_{N}\}$ given the observed data $d$ and model $\model$~\cite{Veitch:2014wba, Ashton:2018jfp}. The 
model-dependent posterior distribution is calculated through Bayes' theorem as,
\begin{equation} \label{eq:bayes}
  p(\boldsymbol{\lambda} \vert d, \model) =  \frac{\Pi(\boldsymbol{\lambda} \vert \model)\, \mathcal{L}(d \vert \boldsymbol{\lambda}, \model)}{\mathcal{Z}},
\end{equation}
where  $\Pi(\boldsymbol{\lambda} \vert \model)$ is the probability of the parameters
$\boldsymbol{\lambda}$ given the model $\model$, otherwise known as the prior;
$\mathcal{L}(d \vert \boldsymbol{\lambda}, \model)$ is the probability of observing the
data given the parameters $\boldsymbol{\lambda}$ and model $\model$, otherwise known
 as the likelihood; and $\mathcal{Z}$ is the probability of observing the data given the model
 $\mathcal{Z} = \int{\Pi(\boldsymbol{\lambda} \vert \model)\, \mathcal{L}(d \vert \boldsymbol{\lambda}, \model)\, d\boldsymbol{\lambda}}$,
otherwise known as the evidence. 

For the case of GW astronomy, the likelihood is known and depends on the specified GW
detector network and individual detector sensitivities~\cite{Veitch:2014wba, Ashton:2018jfp} -- typically characterised by the
power spectral density (PSD) (the square of the ASD). \new{Under the assumption of Gaussian
and stationary noise, t}he likelihood is simply\footnote{\new{The likelihood also includes an additional term describing the noise covariance. Under the assumption that the noise is gaussian and stationary, the noise covariance matrix is the identity matrix and often excluded for simplicity. When this assumption is no longer valid, the noise covariance should be included, see, e.g., Ref.~\cite{Edy:2021par} for details.}}
\begin{equation}
    \mathcal{L}(d | \boldsymbol{\lambda}, \model) = \prod_{k} \exp{\left(-\frac{1}{2}\left(d_{k} - \model(\boldsymbol{\lambda})\, |\, d_{k} - \model(\boldsymbol{\lambda})\right)\right)},
\end{equation}
where $k$ denotes the individual detectors in the network, $(a | b)$ refers to the inner
product between two frequency series $a(f)$ and $b(f)$,
\begin{equation}
    (a | b) = 4\Re \int df\, \frac{a(f)b^{*}(f)}{S_{k}(f)},
\end{equation}
and $S_{k}(f)$ is the PSD of the detector $k$. Despite the likelihood being well known, it
is often not possible to analytically calculate the model dependent posterior distribution.
The reason is because the evidence involves computing the likelihood and prior for all points in
the parameter space. As a result, stochastic sampling techniques, such as Markov-Chain
Monte-Carlo (MCMC)~\cite{metropolis1949monte} and Nested Sampling~\cite{Skilling2004,Skilling:2006}, were developed to return a set of 
independent draws from the unknown posterior distribution; although see e.g. Refs.~\cite{Pankow:2015cra,Lange:2018pyp,Delaunoy:2020zcu,Green:2020hst,Chua:2019wwt,Green:2020dnx,Dax:2021tsq,Gabbard:2019rde,Tiwari:2023mzf,Fairhurst:2023idl} for other 
approaches.

Often we wish to consider a specific configuration of GW detectors. As such, a simulated
GW signal is often injected into real or synthetic GW strain data $n(t)$, or injected into
``zero-noise'' where $n(t)=0\, \forall t$. For either case, the likelihood for a single detector
network\footnote{The likelihood may also be simplified for a multi-detector network. However, we only consider a single detector network for simplicity.}
reduces to,
\begin{equation} \label{eq:posterior_mismatch}
    \mathcal{L}(d | \boldsymbol{\lambda}, \model) \propto A(n) \exp\left(-|h - \model(\boldsymbol{\lambda})|^{2} \right),
\end{equation}
where $|\cdot| = \sqrt{(\cdot| \cdot)}$. The likelihood can be further simplified by introducing
the mismatch: a quantity which characterises how similar two GWs are to
one another. The mismatch between waveforms
$h$ and $\model(\boldsymbol{\lambda})$ ranges between 0 and 1, where 0 implies the 
waveforms are identical (up to an overall amplitude rescaling), and 1 implies the waveforms
are orthogonal. The mismatch is defined as
\begin{equation} \label{eq:mismatch}
    \mismatch = \new{1 -} \max_{dt, d\phi} \frac{(h\, |\, \model(\boldsymbol{\lambda}))}{|h|\, |\model(\boldsymbol{\lambda})|},
\end{equation}
where we maximise over time, $dt$, and phase shifts, $d\phi$. \new{When restricting
attention to a reduced subspace that neglects the phase and time, t}he likelihood takes the simplified form\new{~\cite{Baird:2012cu}}\footnote{\new{It is not necessary to restrict attention to a reduced subspace that neglects the phase and time. If the full space is considered, the match, $1 - \mismatch$, is replaced by the overlap. The overlap follows the same functional form as the match but does not maximise over the phase and time shifts.}},
\begin{equation} \label{eq:simple_like}
    \mathcal{L}(d | \boldsymbol{\lambda}, \model) \propto A(n) \exp\left(-|h|^{2}(1 - \mismatch^{2}) \right).
\end{equation}

For a numerical relativity simulation injected in zero-noise,
Eq.~\eqref{eq:simple_like} implies that the likelihood will peak in the region of the parameter
space where the mismatch between the simulation and $\model(\boldsymbol{\lambda})$
is minimized. For the idealised case, where $\model$ perfectly 
describes numerical relativity, the likelihood will peak at the true parameters of 
the simulated GW signal, $\boldsymbol{\lambda}_{\mathrm{inj}}$. Of course, in reality we expect to observe a bias due to mismodelling; a phenomenon
known as waveform systematics. For this case, the maximum likelihood (and hence the 
minimum mismatch between $h$ and $\model$) will be at the parameters
$\boldsymbol{\lambda}_{\mathrm{max}} \neq \boldsymbol{\lambda}_{\mathrm{inj}}$.
Although in general the maximum likelihood may not peak at the true parameters,
the 90\% credible interval of the posterior distribution may encase
$\boldsymbol{\lambda}_{\mathrm{inj}}$. \new{Likewise, given the dependence of the
prior distribution in Eq.~\eqref{eq:bayes}, the posterior distribution may not peak at
$\boldsymbol{\lambda}_{\mathrm{inj}}$, unless the prior distribution is uniform in all parameters. This may not always be true, especially for derived quantities.}

If we assume the high-SNR limit, where the likelihood is Gaussian,
a conservative limit (see Ref.~\cite{Thompson:2025aaa} for details) for when two waveforms will be distinguishable at 90\% confidence is when
\begin{equation} \label{eq:distinguishability}
	\mismatch \lessapprox  \frac{3.12}{\rho^{2}},
\end{equation}
where $\mismatch$ is the mismatch between the waveforms $h_{1}$ and $h_{2}$ and $\rho$ is the SNR of the signal~\cite{Owen:1995tm,Baird:2012cu}. For the case of a real gravitational-wave signal $h$ and a
model evaluated at the parameters $\boldsymbol{\lambda}$, we can obtain an approximate contour containing 90\% of the posterior distribution by combining Eqs.~\eqref{eq:simple_like} and \eqref{eq:distinguishability}, see e.g. Ref.~\cite{Fairhurst:2023idl}.
Although the 90\% likelihood surface may encase the true parameters, the 
posterior distribution may not owing to shifts caused by the prior.
For a numerical relativity simulation injected into synthetic GW strain data, we expect to 
observe a systematic bias even in the idealised case where the model perfectly describes numerical relativity. 
However, averaging the results from 
many analyses of the same simulation injected into different instances of synthetic GW strain 
data is expected to obtain the same result as a zero-noise injection.

When an ensemble of models exist, $\boldsymbol{\model} = \{\model_{1}, \model_{2}, ...,\model_{i}\}$, individual model-dependent posterior distributions can
be compared and contrasted against the true source properties in order to determine which
model more accurately describes the simulated signal (see Sec.~\ref{Sec:NR_informed_PE} for details about how to combine individual model-dependent posterior distributions to marginalize over model uncertainty). To quantify this a recovery
score for each dimension $\lambda_{j}$ can be calculated, which accounts for the width of the marginalized one-dimensional
posterior as well as the injected value. The recovery score for a given model $\model_{i}$ and simulation $K$ is defined as 
\be
\rat_{K}(\model_{i}):=\f{\sigma_{\lambda_j}(\model_{i})}{C(\lambda_{j}, \model_{i})}\label{eq:ratio_std_to_cost}.
\ee
where $\sigma_\lambda$ is the standard deviation of the one-dimensional marginalized
posterior distribution defined as,
\begin{equation}
    p(\lambda_{j} | d, \model_{i}) = \int p(\boldsymbol{\lambda} | d, \model_{i})\, d\lambda_{1} ... d\lambda_{j - 1} d\lambda_{j + 1}... d\lambda_{N}
\end{equation}
and $C(\lambda_{j}, \model_{i})$ is a cost function given by Ref.~\cite{Knee:2021noc} as
\be
C(\lambda_{j}, \model_{i}) = \int_{-\infty}^\infty p(\lambda_{j} | d, \model_{i})\, (\lambda_{j}-\lambda_{j,\text{inj}})^2 \text{d} \lambda_{j} \label{eq:cost_function}
\ee 
with $\lambda_{j,\text{inj}}$ representing the injected value. If the mean of the
one-dimensional posterior equals $\lambda_{j,\text{inj}}$ then the cost function equals
$\sigma_{\lambda_j}$, otherwise it is greater than it. This assures that
$\rat \in (0,1]$ with
values close to unity highlighting better recovery performance.
For example, assuming a normal distribution with zero mean and standard deviation of $\sigma$, we obtain $\rat \approx\{0.71, 0.45, 0.32\}$ for $\lambda_\text{inj}=\{1,2,3\}\sigma$. Thus, $\rat > 0.7$ indicates robust recovery within $\pm1$ standard
deviation. Note that the particular integral \eqref{eq:cost_function} for the cost function can penalize narrow asymmetric (skewed) posteriors more heavily than wide symmetric posteriors. We will encounter examples
of this in Sec.~\ref{Sec:NR_informed_PE}.

In this work, we perform Bayesian inference on 35 numerical relativity simulations injected
into zero-noise to assess the accuracy of three state of the art precessing GW models. 
Throughout this work, we use a theoretical PSD for Advanced LIGO’s O4 design sensitivity~\cite{O4PSD}, and 
assume a two-detector network consisting of LIGO Hanford (H1) and LIGO Livingston (L1)~\cite{TheLIGOScientific:2014jea}. Although 
numerous packages are now available to stochastically sample the parameter space~\cite{Veitch:2014wba,Ashton:2018jfp,Romero-Shaw:2020owr,Biwer:2018osg}, we use 
the {\texttt{dynesty}} nested sampler \cite{Speagle:2020} via {\texttt{bilby}} \cite{Ashton:2018jfp} and {\texttt{bilby\_pipe}}~\cite{Romero-Shaw:2020owr}. For
all analyses, we use 1000 live points along with the {\texttt{bilby}}-implemented {\texttt{acceptance-walk}} sampling 
algorithm with an average of 60 accepted steps per MCMC. We use sufficiently wide and 
uninformative priors for all parameters. Specifically we use the \texttt{bilby} function
\verb|UniformInComponentsChirpMass| with the range $\Mtot_\text{c} \in [10,40]\Msun$ and 
the \verb|UniformInComponentsMassRatio| function with range $q \in [0.083,1]$
(as well as a constraint on $m_i \in [1,1000]\Msun$).
The luminosity distance prior range is given by $d_\text{L}\in [100,5000]\,$Mpc
via the function \verb|UniformSourceFrame|  using $\Lambda$CDM cosmology
parameterized by Planck 2015 data \cite{Planck:2015fie}.
The spin magnitudes are uniform in the range $\chi_i\in[0,0.99]$ while
the spin tilt angles are uniform in their sines: $\sin\theta_i\in [0,1]$,
and the $\mathbf{J}$-frame inclination angle is uniform in cosine: $\cos{\vartheta_\text{JN,0}}\in [-1,1]$, where
${\vartheta_\text{JN,0}}$ is the angle between $\mathbf{J}$ and $\hat{\mathbf{N}}$ at $f=f_0$.
Other angles are all uniform in their respective ranges.
We consistently integrate the likelihood between $23 - 1024$ Hz, with the injected numerical relativity waveform always generated from $f_0=20$\,Hz to avoid artefacts in the fast Fourier transforms. Although this means that we will be missing power from \emph{e.g.} the $\ell < 5$ higher order multipoles for frequencies $ < 50$ Hz, this was chosen to maximise the power in the dominant quadrupole.

\section{Results I: Recovery Performance of the Waveform Models}\label{Sec:recovery_3models}

As our work presents injection-recovery results of $15$ parameters by three waveform models
for almost three dozen numerical relativity simulations, we calculate and compare recovery
scores for each dimension. We note that the recovery score, as defined in
Eq.~\eqref{eq:ratio_std_to_cost}, has its limitations. Namely, when the injected value lies at 
the edge of the parameter space, as is the case with the recovery of the mass ratio
for the $Q=q=1$ subset of simulations, the recovery score will be $\approx 0.55$ as opposed $\gtrsim 0.7$ for $Q>1$ cases. 
Similarly, if the underlying posterior is heavily skewed, as can happen even without posterior railing,
the recovery score will be reduced.

Another limitation of Eq.~\eqref{eq:ratio_std_to_cost} is that it yields unity whether the posteriors are narrow or wide
as long as the sample mean coincides with the injected value.
To distinguish between such possibilities, we introduce a measure of how sharply peaked the posterior is
via a quantity that we dub the recovery width
\be
\wid := \f{\Sigma_{90}(p(\lambda_{i} | d, \model_{i}) )}{\text{max}(\Pi(\lambda_{i}))-\text{min}(\Pi(\lambda_{i}))} \label{eq:width_ratio}
\ee
with $\Sigma_{90}(p(\lambda_{i} | d, \model_{i}) )$ representing the width of the 90\% credible interval
of the one-dimensional marginalized posterior distribution for the selected parameter
$\lambda_{i}$.
We divide this quantity by the length of the interval for the prior $\Pi(\lambda_{i})$
of each parameter.
Note that given the chirp mass and mass ratio prior ranges of $[10\Msun,40\Msun]$ and $[0.083,1]$, we obtain a rather wide range for the total mass: 
$\Pi(\Mtot) \in [23\Msun,196\Msun]$. As a result, the recovery width values for the total
mass may appear small.
However, we wish to keep the denominator of Eq.~\eqref{eq:width_ratio} the
same for a given parameter as we increase the mass asymmetry of the systems.
This also raises the question of what constitutes a good value for the recovery width
which is not so clear as in the case of the recovery score.
In general, it makes sense to compare values of $\wid$ for the same parameter 
between different models or as the mass ratio is changed, rather than comparing the values of $\wid$ between two different parameters.

\begin{table}
\begin{center}
 \begin{tabular}{lccccccc}
 \hline
\hline
  \texttt{SXS:BBH}& \quad $\Mtot(\Msun)$ & $\ \ \ Q \ \ \ $ & $\ \ a_1\ \ \ $ & $\ \ \ a_2\ \ \ $ & $\ \ \ \theta_1\ \ \ $ & $\ \ \ \theta_2\ \ \ $ & $\ \ \ \Delta\phi\ \ \ $ \\
  \hline
\texttt{0764}	& $	51.5	$ & $	1.0	$ & $	0.80	$ & $	0.80	$ & $	92.9^\circ	$ & $	83.5^\circ $    & 1.53      \\
\texttt{0767}	& $	51.3	$ & $	1.0	$ & $	0.80	$ & $	0.80	$ & $	87.3^\circ	$ & $	92.0^\circ $       &  3.66 \\
\texttt{0838}	& $	51.5	$ & $	1.0	$ & $	0.80	$ & $	0.80	$ & $	93.0^\circ	$ & $	83.5^\circ $      &  1.56 \\
\texttt{0841}	& $	51.3	$ & $	1.0	$ & $	0.80	$ & $	0.80	$ & $	87.1^\circ	$ & $	92.1^\circ $      &  3.68 \\
\texttt{0925}	& $	51.7	$ & $	1.0	$ & $	0.80	$ & $	0.80	$ & $	83.2^\circ	$ & $	92.7^\circ $      &  4.76 \\
\texttt{0935}      & $   51.3 $ & $1.0    $ & $0.80 $ & $0.80	$ & $89.7^\circ	$ & $90.2^\circ$  &       2.60 \\
\texttt{0965}	& $	51.7	$ & $	1.0	$ & $	0.80	$ & $	0.80	$ & $	82.8^\circ	$ & $	91.3^\circ $      &  5.24 \\
\texttt{0982}	& $	51.3	$ & $	1.0	$ & $	0.80	$ & $	0.80	$ & $	93.3^\circ	$ & $	85.0^\circ $      &  2.06 \\
\texttt{1205}	& $	51.4	$ & $	1.0	$ & $	0.85	$ & $	0.85	$ & $	84.4^\circ	$ & $	93.4^\circ $      &  4.19 \\
\texttt{1217}	& $	51.8	$ & $	1.0	$ & $	0.85	$ & $	0.80	$ & $	86.2^\circ	$ & $	86.2^\circ $      &  6.28 \\
\hline
\texttt{0716}	& $	53.6	$ & $	2.0	$ & $	0.80	$ & $	0.80	$ & $	86.2^\circ	$ & $	84.8^\circ $      &  0.06 \\
\texttt{0717}	& $	53.7	$ & $	2.0	$ & $	0.80	$ & $	0.80	$ & $	86.0^\circ	$ & $	96.0^\circ $      &  4.30 \\
\texttt{0812}	& $	53.5	$ & $	2.0	$ & $	0.80	$ & $	0.80	$ & $	89.1^\circ	$ & $	91.3^\circ $      &  3.24 \\
\texttt{0814}	& $	53.9	$ & $	2.0	$ & $	0.80	$ & $	0.80	$ & $	83.9^\circ	$ & $	92.8^\circ $      &  5.36 \\
\texttt{0926}	& $	53.5	$ & $	2.0	$ & $	0.80	$ & $	0.80	$ & $	84.8^\circ	$ & $	95.7^\circ $      &  4.84 \\
\texttt{0936}	& $	53.5	$ & $	2.0	$ & $	0.80	$ & $	0.80	$ & $	90.9^\circ	$ & $	86.9^\circ $      &  2.69 \\
\texttt{0966}	& $	53.1	$ & $	2.0	$ & $	0.80	$ & $	0.80	$ & $	85.5^\circ	$ & $	94.5^\circ $      &  5.23 \\
\texttt{0976}	& $	53.4	$ & $	2.0	$ & $	0.80	$ & $	0.80	$ & $	87.8^\circ	$ & $	93.9^\circ $      &  3.73 \\
\texttt{1197}	& $	52.7	$ & $	2.0	$ & $	0.85	$ & $	0.85	$ & $	92.5^\circ	$ & $	83.1^\circ $      &  2.04 \\
\texttt{1200}	& $	53.6	$ & $	2.0	$ & $	0.85	$ & $	0.85	$ & $	85.9^\circ	$ & $	84.3^\circ $      &  0.06 \\
\hline
 \texttt{1916}	& $	58.3	$ & $	4.0	$ & $	0.80	$ & $	0.80	$ & $	87.5^\circ	$ & $	98.7^\circ$     &  4.18 \\
\texttt{1921}	& $	59.0	$ & $	4.0	$ & $	0.80	$ & $	0.80	$ & $	87.6^\circ	$ & $	92.8^\circ $      &  3.28 \\
\texttt{1922}	& $	58.3	$ & $	4.0	$ & $	0.80	$ & $	0.80	$ & $	88.9^\circ	$ & $	84.3^\circ $      &  2.36 \\
\texttt{1923}	& $	59.1	$ & $	4.0	$ & $	0.80	$ & $	0.80	$ & $	83.4^\circ	$ & $	88.4^\circ $      &  5.97 \\
\texttt{2000}	& $	59.8	$ & $	4.0	$ & $	0.80	$ & $	0.80	$ & $	84.0^\circ	$ & $	97.7^\circ $       & 5.13 \\
\texttt{2004}	& $	58.2	$ & $	4.0	$ & $	0.80	$ & $	0.80	$ & $	88.4^\circ	$ & $	96.3^\circ $      &  3.77 \\
\texttt{2070}	& $	58.4	$ & $	4.0	$ & $	0.80	$ & $	0.80	$ & $	89.5^\circ	$ & $	86.3^\circ $      &  2.61 \\
\texttt{2074}	& $	59.1	$ & $	4.0	$ & $	0.80	$ & $	0.80	$ & $	84.7^\circ	$ & $	83.1^\circ $      &  0.23 \\
\texttt{2075}	& $	58.6	$ & $	4.0	$ & $	0.80	$ & $	0.80	$ & $	86.8^\circ	$ & $	95.2^\circ $      &  5.53 \\
\texttt{2079}	& $	58.9	$ & $	4.0	$ & $	0.80	$ & $	0.80	$ & $	87.3^\circ	$ & $	98.7^\circ $      &  4.07 \\   
\hline
\hline
  \end{tabular}
  \caption{The most relevant intrinsic parameters for the 30 \SXS{} binary black hole simulations used throughout this work. Column one lists the \SXS{} simulation code
  with columns two and three displaying the total system mass (in $\Msun$) and the
  mass ratio, respectively. Columns four and five give the total dimensionless spin magnitudes
  for the black holes while columns six, seven and eight list the spin tilt angles and the relative azimuthal separation of the spin vectors.}
  \label{tab:SXS_sim_params}
\end{center}
\end{table}

To further aid in comparing model performance, we additionally introduce
\be
\rat_\text{av}(\model_{i}) = \frac{1}{N} \sum_{K} \rat_\texttt{K} (\model_{i}) \label{eq:av_ratio},
\ee
where \texttt{K} represents the set of $N$ numerical simulations to average over. The important
thing to keep in mind when looking at the values of $\rat$ and $\wid$ is their magnitude for a given model
relative to another model or for a given subset of simulations, e.g., $Q=1$, vs. another subset, e.g., $Q=2$.

\subsection{Cases with mass ratio between $1:1$ and $4:1$}\label{sec:Qle4_injections}

We analysed a subset of 30 numerical relativity simulations produced by the \SXS{}
collaboration with mass ratios between $1:1$ and $4:1$. We selected these simulations according to the following criteria. We start with the requirement that 
the number of orbital cycles is $> 18$, residual NR eccentricity less than $ 10^{-3}$, $\vert a_i \vert > 0.79$ and $ 1 < \cos^{-1}\theta_i < 2$. This yields 18 simulations with $Q<1.2$, 57 with $1.2 < Q<2.2$ and 24 with $Q>2.2$. We reduce the size of the $1.2<Q<2.2$ portion by limiting the corresponding tilt angles to be within $\pm1\%$ of the range of the corresponding $Q<1.2$ angles while also
restricting the range of $\Delta\phi$ to be $5\%-95\%$ of the range of the $Q<1.2$ values. 
This yields 10 cases with $1.99958\le Q\le 2.00011$. We obtain 10 cases with $3.99911\le Q \le 4.00046$ in a similar fashion, but without needing to restrict $\Delta\phi$.
Finally, we relax the $\Delta\phi$ bound enough to obtain 10 cases with $1.00001 \le Q\le 1.00006$. These steps produce a dataset that has very similar parameters in the dimensionless
spin space, but this is deliberate as we want to focus on keeping the strength of precession
more or less fixed while we increase the mass asymmetry.
Finally, we checked that none of our chosen simulations have been listed as deprecated in the updated \SXS{}
catalog \cite{Scheel:2025jct}.

The most relevant intrinsic parameters for the resulting 30 BBHs are given
in Table~\ref{tab:SXS_sim_params}. The values for the detector-frame binary total mass are chosen such that the detector-frame reference GW frequency of the \SXS{} simulation is 20\,Hz.
Note that all dimensionless spin magnitudes are set to the same high value, i.e., $a_1=a_2=0.8$, but the spin tilt angles are slightly different in each case though they are all between $80^\circ$ and $ 100^\circ$, resulting in systems with rather strong precession, i.e., $\chip > 0.75$.
The values of $\{\iota, \alpha, \delta, \psi, d_\text{L}\}$ for each of the 30 BBHs are chosen such that the total network SNR, $\rho_\text{tot}$, is 40, with the precession SNR \cite{Fairhurst:2019vut}, $\rho_{\mathrm{p}}$, equalling 10.
The specific values of the extrinsic parameters can be found in Table~\ref{tab:SXS_extrinsic_params} in App.~\ref{Sec:App_A}
where it can be seen that many different combinations of extrinsic parameters can yield $\rho_\text{p}=0.25\rho_\text{tot}$.

\begin{figure*}[tp!]
    \centering
    \includegraphics[width=0.99\textwidth]{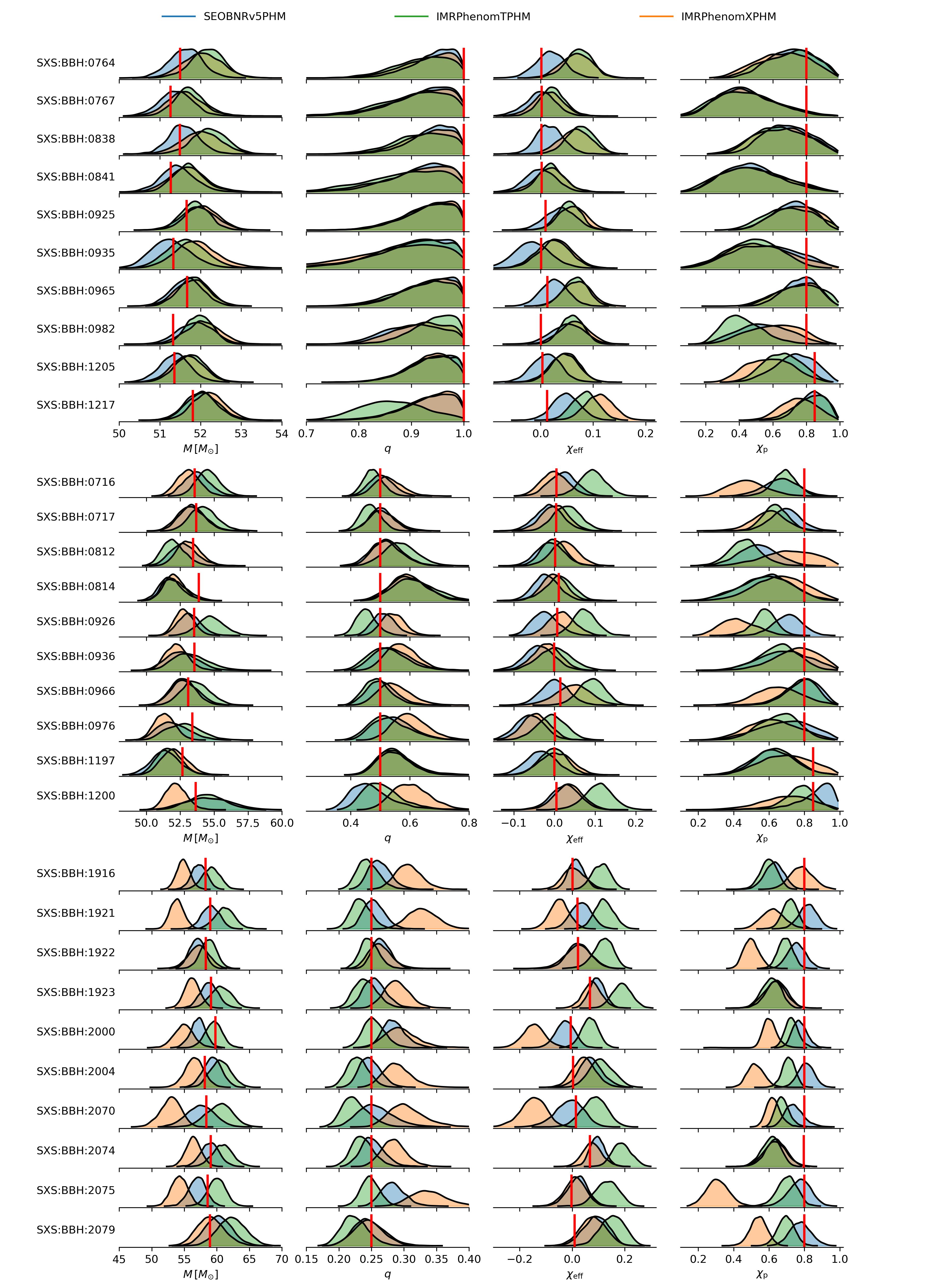}
\caption{One-dimensional marginalized posterior distributions obtained for the inferred total detector-frame mass (first column), mass ratio (second column), effective parallel (third column) and effective perpendicular spins (fourth column) for the 30 \SXS{} binary black hole simulations used throughout this work. The red vertical lines indicate the true values.}
\label{fig:three_model_posteriors}
\end{figure*}

In Fig.~\ref{fig:three_model_posteriors}, we present the one-dimensional marginalized posterior distributions 
for $\{\Mtot, q, \chieff, \chip \}$ as recovered by \SEOB, \TPHM{} and \XPHM.
The figure contains 30 rows divided into subsets of 10 by mass ratio and four columns, one
for each parameter. The top (bottom) 10 rows show the $q=1\, (1/4)$ results with the middle 10 displaying
the $q=1/2$ posteriors. The number for each \SXS{} simulation is at the left end of each row.
We summarize our key findings below.

 %==============================================================
%    FIG: pulse plots: 3 models, 30 sims, 4 parameters
%==============================================================
%==============================================================

\subsubsection{Mass ratio $1:1$ systems}\label{sec:Q1_subset}

\begin{table*}
\begin{center}
 \begin{tabular}{lccc|ccc|ccc|ccc}
 \hline
\hline
$\Theta$	 & 	$\rat_\text{av}^\textsc{S} $	 & 	$\rat_\text{av}^\textsc{T} $	 & 	$\rat_\text{av}^\textsc{X} $	 & 	$\rat_\text{av}^\textsc{S}[Q=1] $	 & 	$\rat_\text{av}^\textsc{T}[Q=1] $	 & 	$\rat_\text{av}^\textsc{X}[Q=1] $	 & 	$\rat_\text{av}^\textsc{S}[Q=2] $	 & 	$\rat_\text{av}^\textsc{T}[Q=2] $	 & 	$\rat_\text{av}^\textsc{X}[Q=2] $	 & 	$\rat_\text{av}^\textsc{S}[Q=4] $	 & 	$\rat_\text{av}^\textsc{T}[Q=4] $	 & 	$\rat_\text{av}^\textsc{X}[Q=4] $	\\
\hline
$ \Mtot $	 & 	0.834	 & 	0.723	 & 	0.625	 & 	0.917	 & 	0.699	 & 	0.733	 & 	0.793	 & 	0.740	 & 	0.686	 & 	0.793	 & 	0.730	 & 	0.456	\\
$ q $	 & 	0.756	 & 	0.708	 & 	0.593	 & 	0.555	 & 	0.531	 & 	0.556	 & 	0.881	 & 	0.783	 & 	0.712	 & 	0.831	 & 	0.810	 & 	0.512	\\
$ \chi_\text{eff} $	 & 	0.807	 & 	0.532	 & 	0.718	 & 	0.834	 & 	0.518	 & 	0.559	 & 	0.786	 & 	0.720	 & 	0.872	 & 	0.803	 & 	0.358	 & 	0.724	\\
$ \chi_\text{p} $	 & 	0.692	 & 	0.512	 & 	0.521	 & 	0.698	 & 	0.634	 & 	0.645	 & 	0.676	 & 	0.538	 & 	0.607	 & 	0.702	 & 	0.365	 & 	0.310	\\
\hline
\hline
\end{tabular}
\caption{The recovery score \eqref{eq:ratio_std_to_cost} 
of each waveform model for the parameters listed in column one.
Columns two to four list the averages over the entire set of 30 simulations for \SEOB, \TPHM, \XPHM,
respectively labeled as \textsc{S, T, X} here. The remaining columns show the average of the same quantity
over the size-10 $Q=1,2,4$ subsets. A value of $\rat >0.7$ indicates a recovery of the
injected parameter within one standard deviation of the mean of the posterior distribution.
Note that the lower recovery score for the mass ratio $q$ for the $q=1$ subset is due to the fact that the injected value
lies at the boundary of the domain hence causing the posteriors to rail as can be seen in Fig.~\ref{fig:three_model_posteriors}. The closer the recovery score to unity, the better.}\label{tab:recovery_scores}
\end{center}
\end{table*}

 For the $Q=q=1$ simulations, all three models recover the detector-frame total mass within their respective 90\% credible
 intervals (CIs), with \SEOB's posteriors generally peaking closer to the injected values. 
Similarly, the models recover the injected value of the mass ratio, i.e., $q=1$ well modulo the usual issues with having
the injected value at the boundary of the parameter space.
Interestingly, for \SXS\texttt{:\,1217} \TPHM's posteriors completely miss the true value. 
However, because \TPHM{} severely misidentifies the mass ratio of the binary, while inferring $\Mtot$
well, \TPHM{} obtains biased estimates for the components masses $m_{1}$ and $m_{2}$.
It is not clear why the model behaves this way for this particular simulation.
We checked the signal power in the subdominant multipoles~\cite{Mills:2020thr}, but found that $\rho_{44}\approx 1.5$ with other multipole SNRs less than one.
As a further check, we re-injected this simulation with the \sxsSim{1205}'s extrinsic parameters (ensuring a fixed total SNR, but varying precession SNR) and
recovered posteriors that are consistent, albeit broader, with the injection.
We also conducted the converse experiment, i.e., re-injecting \sxsSim{1205} with \sxsSim{1217}'s extrinsic parameters.
This time, we observed the expected posterior railing against $q=1$ as opposed to what Fig.~\ref{fig:three_model_posteriors} shows for \sxsSim{1217}.
As a final check, we re-performed the PE analysis with more aggressive settings to ensure that the
posterior distribution is converged
(e.g., twice as many live points, see the config file in our public data release). We found that the resulting posteriors are statistically indistinguishable.
Moreover, when plotting the 90\% credible interval for the reconstructed [whitened] waveforms
we find that the recovered \TPHM{} posterior overlaps better with the injected \SXS{} simulation
than when generating the injection with \TPHM{} ({i.e.,} when passing the injected
parameters to the \TPHM{} waveform model). Additionally, the maximum-likelihood
\TPHM{} waveform yields a marginally larger match of 0.976 to the injected \SXS{} simulation, compared to
0.975 when matching the injection generated with \TPHM{} against the
injected \SXS{} simulation.
Therefore, we conclude that this unexpected recovery result by \TPHM{} is due
to specific combination of intrinsic and extrinsic parameters exposing a well-known
fact of GW data analysis: the intrinsic parameters for the maximum-likelihood waveform need not agree well with the injected intrinsic parameters.
This is a generic feature of the stochastic nature of the data analysis methods, and not limited to just one waveform model.

 The models show more variability in their recovery of $\chieff$ with \SEOB{} always recovering the true source properties within
 its 90\% CI, and peaking close to the true value half the time.
 Interestingly, \TPHM{} and \XPHM{} seem to recover $\chieff$ worse when \SEOB{} recovers it near its peak
 as can be seen, e.g., in the cases of \SXS\texttt{:\,0764, 0838, 0965}. These simulations all share very similar
 values for $\theta_1, \theta_2$ that yield $\vert\chieff\vert \lesssim 0.05$. 
 It is also noteworthy that \SEOB's performance degrades for a similar configuration, \SXS\texttt{:\,0982},
 compared to, e.g., \SXS\texttt{:\,0838}. This turns out to be caused by the railing of \SEOB's posteriors for
 $a_1$ against zero, thus missing the true value of 0.8 by more than three standard deviations (SDs or $\sigma$'s).
 Therefore, the resulting \SEOB{} posterior favours $\chieff> 0$ at 90\% confidence. 
 Nonetheless, this is marginally better than the output of \TPHM{} and \XPHM.
 
 \begin{table*}[t!]
\begin{center}
 \begin{tabular}{lccc|ccc|ccc|ccc}
 \hline
\hline
$\Theta$	 & 	$\wid_\text{av}^\textsc{S} $	 & 	$\wid_\text{av}^\textsc{T} $	 & 	$\wid_\text{av}^\textsc{X} $	 & 	$\wid_\text{av}^\textsc{S}[Q=1] $	 & 	$\wid_\text{av}^\textsc{T}[Q=1] $	 & 	$\wid_\text{av}^\textsc{X}[Q=1] $	 & 	$\wid_\text{av}^\textsc{S}[Q=2] $	 & 	$\wid_\text{av}^\textsc{T}[Q=2] $	 & 	$\wid_\text{av}^\textsc{X}[Q=2] $	 & 	$\wid_\text{av}^\textsc{S}[Q=4] $	 & 	$\wid_\text{av}^\textsc{T}[Q=4] $	 & 	$\wid_\text{av}^\textsc{X}[Q=4] $	\\
\hline
$ \Mtot $	 & 	0.00326	 & 	0.00322	 & 	0.00354	 & 	0.00330	 & 	0.00306	 & 	0.00349	 & 	0.00343	 & 	0.00351	 & 	0.00386	 & 	0.00304	 & 	0.00309	 & 	0.00327	\\
$ q $	 & 	0.614	 & 	0.607	 & 	0.667	 & 	0.623	 & 	0.577	 & 	0.658	 & 	0.647	 & 	0.662	 & 	0.727	 & 	0.573	 & 	0.583	 & 	0.616	\\
$ \chi_\text{eff} $	 & 	0.282	 & 	0.278	 & 	0.306	 & 	0.286	 & 	0.265	 & 	0.302	 & 	0.297	 & 	0.304	 & 	0.334	 & 	0.263	 & 	0.267	 & 	0.283	\\
$ \chi_\text{p} $	 & 	0.563	 & 	0.557	 & 	0.612	 & 	0.571	 & 	0.529	 & 	0.603	 & 	0.593	 & 	0.607	 & 	0.667	 & 	0.525	 & 	0.534	 & 	0.565	\\
\hline
\hline
\end{tabular}
\caption{Same as Table~\ref{tab:recovery_scores}, but for the recovery width $\wid$.
Smaller recovery widths indicate narrower posterior distributions.
As the denominator of Eq.~\eqref{eq:width_ratio} yields much
larger values for $\Mtot$ than $\{q,\chieff,\chip\}$ relative to the numerator, the
resulting recovery widths for $\Mtot$ may be interpreted to be artificially small.
An alternative is replacing the denominator with $\text{mean}(p(\Mtot))$ which
yields values of $\wid\approx \ord(10^{-2})$ for the total mass.}\label{tab:width_ratio}
\end{center}
\end{table*}

  The models' recovery performance for $\chip$ is rather complicated as this depends on the recovery of
 $\{m_{1}, m_{2}, a_1,a_2,\theta_1,\theta_2\}$.
 Let us first focus on the first seven $Q=1$ cases, i.e., \SXS\text{:\,0764} to \texttt{0965} for which all three models behave the same way.
 We see that they seem to alternate between either recovering the injected values close to their peaks or
 underestimating it by one to two SDs. 
 A deeper look reveals that all three models recover $a_1$ and $a_2$ poorly for \texttt{767} and \texttt{841},
 but produce narrow posteriors for $\theta_1,\theta_2$. This means that $\chieff \approx 0$ can still be well recovered
 while $\chip$ recovery is poor.
 The only exception seems to be \SXS\text{:\,0935} where the posteriors for $\{a_1,a_2,\theta_1,\theta_2\}$
 contain the injected values within their 90\% CIs. However, the posteriors for $\theta_1,\theta_2$ are much
 broader in this case than the other six cases, hence the poor $\chip$ recovery.
 We have also noticed that for this particular case \TPHM{} yields posteriors that strongly peak at 
 $\theta_1\approx 2.5, \theta_2\approx 0.5$, thus further skewing its $\chip$ posterior\footnote{The 1D posteriors for all the system parameters can be seen on our aforementioned \texttt{git} repository.}
\TPHM{} additionally produces divergent results for \SXS\text{:\,0982}, a simulation where all models struggle with recovery. Interestingly,
\XPHM{} marginally outperforms \SEOB{} for this case.
A detailed investigation reveals that \TPHM{} poorly recovers each of the four parameters in the set $\{a_1,a_2,\theta_1,\theta_2\}$
with \SEOB{} and \XPHM{} recovering $\{a_2,\theta_1,\theta_2\}$ reasonably well.
\SXS\text{:\,0982} is most similar to \SXS\text{:\,0838} when it comes to its intrinsic parameters so it is puzzling that
\SEOB{} should perform so differently between these two simulations whereas \TPHM{} and \XPHM{} behave somewhat
consistently.

\new{Recalling that the models considered here obey the multipolar symmetry \eqref{eq:h_coprec_symmetry},
an explanation for the poor $\chip$ recovery for at least
\sxsSim{0767, 0841} and possibly \sxsSim{0935} is the fact that these simulations
have their values of $\Delta\phi$ closest to $\pi$ as listed in Table~\ref{tab:SXS_sim_params}.
Mass ratio $1:1$ systems with spins purely in the orbital plane pointing in the opposite directions, i.e., $\Delta\phi=\pi$, are well known to be in the so-called superkick configuration \cite{Gonzalez:2007hi, Campanelli:2007cga, Tichy:2007hk},
having the largest amount of multipole asymmetry.
}

We can quantify some of the above statements further via the values for the recovery score $\rat$ and {recovery width} $\wid$
presented in Tables~\ref{tab:recovery_scores} and \ref{tab:width_ratio}, specifically columns five through seven, showing
the averaged values for the $Q=1$ subset of simulations. We see that \SEOB's average $\rat$ exceeds 0.9 for $\Mtot$
corroborating what Fig.~\ref{fig:three_model_posteriors} shows.
For the mass ratio $q$, we obtain very similar values of $\rat_\text{av}$ from all three models with \TPHM's score slightly worse because of the aforementioned issues related to \sxsSim{1217}.
As \TPHM{} and \XPHM{} $\chieff$ posteriors are very similar, the resulting $\rat_\text{av}$'s are in the range $0.5$-$0.6$
with \SEOB's exceeding $0.8$, indicating very good recovery performance.
Finally, since all three models produce very similar $\chip$ posteriors for seven of the 10 $Q=1$ simulations, the resulting
$\rat_\text{av}$ are also similar with some differences due to model performance for \sxsSim{0982}, \texttt{1205} and \texttt{1217}.

The {recovery widths} of the model posteriors are in general quite comparable for the $Q=1$ subset with \TPHM{} (\XPHM) producing the narrowest (widest)
posteriors for all four parameters considered in Table~\ref{tab:width_ratio}.
However, recall that a small recovery width is only meaningful when coupled with
a good recovery score. For example, though \TPHM{} has the narrowest
posteriors for $\chieff$, it yields $\rat_\text{av}=0.518$, much lower than \SEOB's
value of $0.834$.

\subsubsection{Mass ratio $2:1$ cases}\label{sec:Q2_subset}

Turning our attention to the  $Q=2\ (q=1/2)$ subset,
we return to Fig.~\ref{fig:three_model_posteriors} but now focusing on the middle 10 panels.
In general, the models recover $\Mtot$ and $q$ within their 90\% CIs with the exception of \sxsSim{0814} 
where all three models fail to recover these parameters.
The intrinsic spin parameters of this system are most similar to the $Q=1$ \sxsSim{0925} case, where we had recorded
robust recovery of $\{\Mtot,q\}$ by all three models. Therefore, it may be another case of the extrinsic parameters
conspiring to make this simulation more challenging.
Indeed, we find that when we re-inject \sxsSim{814} with \sxsSim{925}'s extrinsic parameters (again ensuring the same network SNR but varying precession SNR) for another recovery run with \XPHM, the resulting posteriors for
$q$ shift close enough to the injected value $q=1/2$ that the recovery is now within
$1\sigma$, albeit with broader posteriors than before.
More quantitatively, the $q$ recovery score jumps from $\rat=0.50$ to $0.78$,
and from $0.38$ to $0.66$ for $\Mtot$.

Beside \sxsSim{0814}, \XPHM{} also fails to recover $\{\Mtot,q\}$ within its 90\% CI for \sxsSim{0976} and \sxsSim{1200}.
It significantly underestimates $\Mtot$ while overestimating $q$. 
This is not unusual as the best recovered parameter is often the chirp mass whose
[marginalized] 2D posteriors form a narrow banana-shaped region in the $m_1$-$m_2$ space with very similar likelihoods along the ``banana'' (for high mass, short duration simulations, the total mass is often the best recovered parameter due analyses only being sensitive to the merger and ringdown portions of the signal). 
Therefore, a model recovering $\cal{M}_\text{c}$ within $\pm1\sigma$ can yield $\{M,q\}$ posteriors due to an under(over)estimation of $m_1$ ($m_2$) such as the ones
mentioned here.

The values in Table~\ref{tab:recovery_scores} corroborate these findings with \SEOB{} yielding the best recovery
scores for both $\Mtot$ and $q$, followed by \TPHM{} then \XPHM.
\SEOB{} also produces the narrowest $\Mtot,q$ posteriors with \TPHM{} the broadest as given by the recovery widths in Table~\ref{tab:width_ratio}.

The recovery of $\chieff$ is in general quite robust among the models. 
\XPHM{} has the highest average recovery score which can also be seen in Fig.~\ref{fig:three_model_posteriors}.
\SEOB's recovery is similar to \XPHM's, but ever so slightly worse in some cases, thus its recovery score being
about 10\% lower.
\TPHM's performance is the least consistent with the model sometimes recovering very well with $\rat\approx 1$,
but also not recovering the injections within the 90\% CI for four of the 10 cases.

It is difficult to draw any general conclusions about the models' recovery performance for spin precession.
What is clear from Fig.~\ref{fig:three_model_posteriors}, even at a mass ratio of $2:1$ and a precession SNR of $10$,
is that all three models struggle to produce a consistently faithful recovery of the injection.
It also seems that time-domain \SEOB{} and \TPHM{} models behave more similarly to each other than either to the frequency-domain \XPHM{} model.
A detailed investigation of the recovery of $\{a_1,a_2,\theta_1,\theta_2\}$ reveals, e.g., that 
\XPHM's performance for \sxsSim{0716} and \texttt{0926} is due to its underestimating $a_1$ by $2\sigma$
to $3\sigma$. \TPHM{} behaves the same way for \sxsSim{0812} and \texttt{0926}.
\SEOB's $a_1$ recovery similarly makes it underestimate $\chip$ by $2\sigma$ for \sxsSim{0812}.
For these three simulations, \SEOB{} and \TPHM{} recover $\theta_1$ robustly, as does \XPHM{} except for
\sxsSim{0716}.
\TPHM{} even seems to recover $\theta_2$ very well with the injected values recovered within $\pm \sigma$.
So the underlying issue, as was the case with the relevant $Q=1$ subset, is the unreliable inference of $a_1$ for certain cases.

The recovery scores of the models for $\chip$, given in Table~\ref{tab:recovery_scores}, quantitatively affirm
the above statements. First, of the four parameters presented in the Table for $Q=2$, 
the $\chip$ recovery scores are significantly lower than the rest for all models though
\SEOB's average recovery score of $0.676$ is high enough to be considered reliable.
Similarly, the recovery widths $\wid$ for the $\chip$ posteriors are roughly twice those of $\chieff$ implying much broader posteriors.  This is expected as spin components perpendicular to $\mathbf{L}$ are subdominant to parallel components in the waveform phase by half a post-Newtonian order \cite{Apostolatos:1994mx, Arun:2008kb}.

\subsubsection{Mass ratio $4:1$ systems}\label{sec:Q4_subset}

A quick glance to the bottom 10 rows of Fig.~\ref{fig:three_model_posteriors} reveals that the posteriors from different models
now overlap much less than they did for the $Q=1,2$ cases, indicating increased
model disagreement for higher-mass ratio configurations.
For $\Mtot$ and $q$, \SEOB{} yields the best recovery performance
with $\rat_\text{av}=0.793, 0.831$, respectively.
The mass ratio recovery is especially impressive with \SEOB{}
capturing the injection within $1\sigma$ for eight out of the 10 cases.
\TPHM's performance is nearly comparable, reflected by its corresponding recovery
scores being less than 10\% lower than \SEOB's.
Additionally, \TPHM{} recovers better scores than \SEOB{} for two cases,
\sxsSim{2000} and \sxsSim{2075}, where \SEOB's recovery scores are the lowest.
The recovery widths of \SEOB{} and \TPHM{} are also quite close for this mass ratio
as can be seen from Table.~\ref{tab:width_ratio}.
\XPHM's recovery performance of $\{\Mtot,q\}$ degrades significantly at this mass ratio with $\Mtot\ (q)$ under(over)estimated by more than $2\sigma$ in six cases, which is also
reflected by the average recovery scores of $0.456\ (0.512)$ given in the last column of Table~\ref{tab:recovery_scores}.

Next, focusing on the recovery of the effective spins, we see that 
\SEOB's $\chieff$ recovery is very robust with $\rat_\text{av}=0.803$.
\XPHM{} also recovers $\chieff$ 
well with $\rat_\text{av}=0.724$
whereas \TPHM's recovery is rather poor. We think this may be
due to our specific choice of $Q\le 4$ simulations, all of which have
$\vert\chieff\vert < 0.02$. 
This poor $\chieff$ recovery performance of \TPHM{} was already observed in Ref.~\cite{MacUilliam:2024oif} for the $\chieff=0.001$ injection \sxsSim{0050}.
For non-negligible values of $\chieff$, \TPHM{} should recover $\chieff$ well as can be seen,
e.g., in Fig.~4 of Ref.~\cite{Estelles:2021gvs}.
This is partly supported by the results of our $Q=8$ injections (Sec.~\ref{sec:Q8_injections}): 
\TPHM{} infers $\chieff$ without biases for two out of three $\vert\chieff\vert > 0.17$ cases.

As for the recovery of the effective precession spin, we observe that \SEOB{} significantly outperforms
\TPHM{} and \XPHM{} with a $Q=4$ average recovery score of $0.702$ vs. $(<0.4)$ for the other
two models.
However, \SEOB{} does not recover $\chip$ well for \SXS\texttt{: 1916, 1923, 2074}. 
In fact, none of the models can recover $\chip$ within $2\sigma$ for the \sxsSim{1923}, \texttt{2074} injections,
though \SEOB{} and \XPHM{} do recover $\chieff$. These two cases have very similar intrinsic and extrinsic 
parameters with the exception of the sky position and polarization angles.
We therefore recommend that future waveform models analyse the \sxsSim{1923} and \texttt{2074} simulations to assess their improved accuracy.
What is more puzzling is the case of \sxsSim{1916} which has intrinsic parameters very similar to \sxsSim{2079} for
which \SEOB{} recovers $\chip$ well (but not \TPHM{} or \XPHM).
This turns out to be due to the $2$-$\sigma$ biased recovery of $a_1$ for this case, which does not occur for \sxsSim{2079}.
Given the similarity of the intrinsic parameters, we suspect once again that particular combinations of the extrinsic parameters cause \SEOB{} to underestimate $\chip$ for \sxsSim{1916} though the $\chieff$ recovery remains robust.
After all, given that the parameter space is 15-dimensional, many combinations of parameters can result in waveforms
that match the data very well.

\new{A potential explanation for \SEOB's deteriorated $\chip$ recovery for \sxsSim{1923, 2070, 2074} may once again be the multipole symmetry \eqref{eq:h_coprec_symmetry}.
Although we do not systematically explore the consequences of this symmetry on the
recovery performance of \SEOB{} (or the other models), studies have been conducted
using \NRsur{} and \SEOB{} \cite{Kolitsidou:2024vub, Estelles:2025zah}}.
%Ref.~\cite{Kolitsidou:2024vub} re-analyzes GW200129 with a version of
%\textsc{NRSur7dq4} that has its multipoles symmetrized and shows that
%the symmetrized model can yield parameter posteriors that are quite different
%(see their Table I).
\new{Especially relevant here is the injection-recovery of \sxsSim{2070} by Ref.~\cite{Estelles:2025zah}
using the same version of \SEOB{} as in here and an improved version that incorporates
mode asymmetry for the $\ell=m\le 4$ multipoles.
In particular, they find that the $\chip$ posteriors shift by more than $3\sigma$ from
overestimating to underestimating the injected value when the multipole symmetry
is removed. Though the ``asymmetric'' \SEOB{} model shows no improvement for this
particular case, it does so for two other injections and a re-analysis of GW200129 \cite{Estelles:2025zah}.}

\new{To supplement their findings, we recovered the same injection with the model
\texttt{IMRPhenomXO4a} which has the $(2,2)\to(2,-2)$ multipole asymmetry.
Interestingly, we find that, as shown in Fig.~\ref{fig:posterior_comparison_XO4a_with_XPHM}, the resulting $\chip$ posteriors form a narrow distribution
strongly peaking slightly father away from the injected value than the mode-symmetric \SEOB. Perhaps, this is due to the fact that the \textsc{LALS}imulation version of \texttt{IMRPhenomXO4a} that we use is missing the phase offset fix implemented in Ref.~\cite{Mielke:2024kya}.
%The obvious thing to try here is to re-recover this injection with this improved
%version of \texttt{IMRPhenomXO4a}.
}

For the same set of $Q=4$ simulations, 
\XPHM{} severely underestimates $\chip$, by more than three standard deviations in some cases,
with the exception of \SXS\texttt{: 1916} which it recovers very well.
This performance is reflected in its average recovery score of $0.310$.
Though \TPHM's performance seems marginally better based on its recovery score of $0.365$,
Fig.~\ref{fig:three_model_posteriors} reveals that \TPHM's posteriors underestimate $\chip$
much less severely than \XPHM's do. The average recovery scores are comparable because of \XPHM's
excellent recovery of $\chip$ ($\rat\approx 1$) for \sxsSim{1916}, which skews its average
from roughly $0.2$ to above $0.3$.

\subsubsection{Summary of findings for mass ratios $< 4: 1$ simulations}

%==============================================================
%    FIG: pulse plots: 3 models, 30 sims, 4 parameters
%==============================================================
\begin{figure*}[t!]
    \centering
    \includegraphics[width=0.99\textwidth]{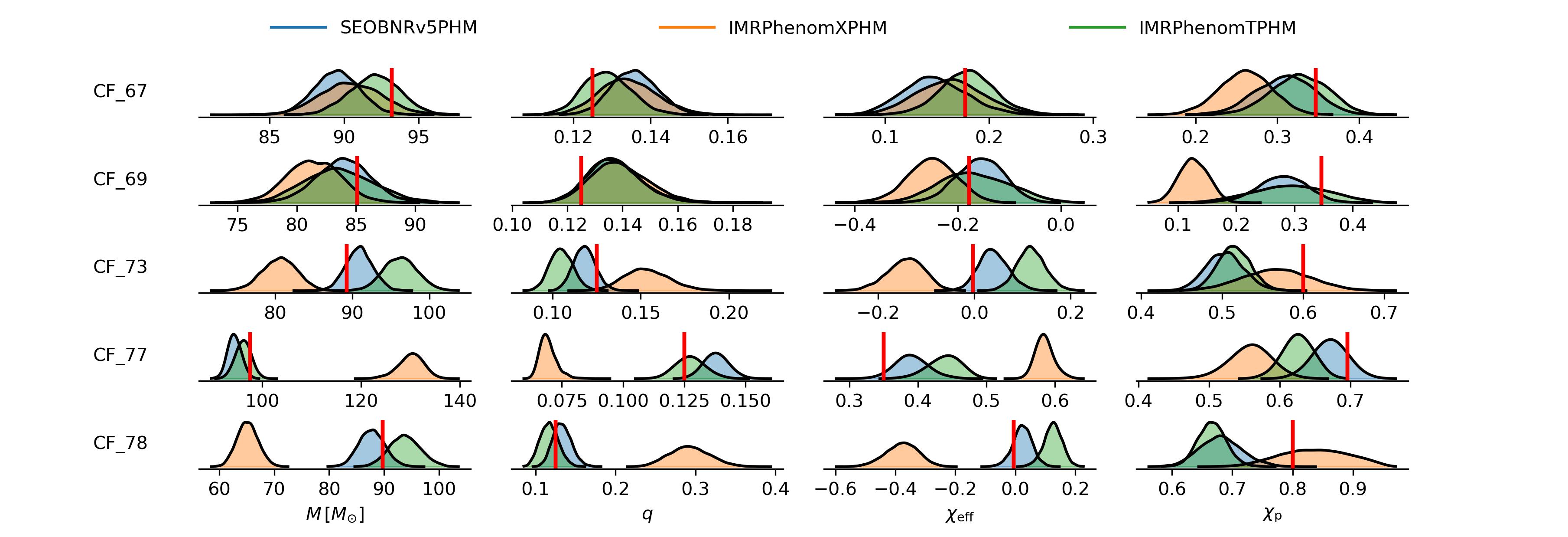}
\caption{Same as Fig.~\ref{fig:three_model_posteriors} but for the five single-spin $Q=8$ \texttt{BAM} 
binary black hole simulations of Sec.~\ref{sec:Q8_injections}. 
The red vertical lines indicate the true values.}
\label{fig:three_model_posteriors_BAM}
\end{figure*}
%==============================================================

\begin{table}
\begin{center}
 \begin{tabular}{lccccc}
 \hline
\hline
 \texttt{CF\_}& \quad $\Mtot(\Msun)$ & $\ \ \ Q \ \ \ $ & $\ \ a_1\ \ \ $ & $\ \ \ \theta_1\ \ \ $  \\
  \hline
\texttt{67}	& $93.2$ &  $	8.0	$ & $0.40$  & $59.9^\circ$ \\
\texttt{69}	& $85.1$ &  $	8.0	$ & $0.40$   & $119.8^\circ$\\
\texttt{73}	& $ 89.3$ &  $	8.0	$ & $0.60$  & $89.8^\circ$  \\
\texttt{77}	& $97.5 $ &  $	8.0	$ & $0.80$   & $59.7^\circ$\\
\texttt{78}	& $89.7 $ &  $	8.0	$ & $0.80$   & $89.6^\circ$\\
\hline
\hline
  \end{tabular}
  \caption{The most relevant intrinsic parameters for the five \texttt{BAM} binary black hole simulations used throughout this work. The columns represent the same quantities as in Table~\ref{tab:SXS_sim_params}. Note that these are all single-spin simulations, i.e., $a_2=0$.}
  \label{tab:BAM_sim_params}
\end{center}
\end{table}

In summary, for all 30 simulations considered here, we can confidently state that \SEOB{} recovers
the set of parameters $\{\Mtot,q, \chieff,\chip\}$ well as indicated by the overall average recovery
scores given in columns 2 to 5 of Table~\ref{tab:recovery_scores}.
Its $\rat_\text{av}=0.692$ for $\chip$ translates to 
a $\pm1$ standard deviation recovery of the effective precession spin from dozens of synthetic GW events with a wide range of extrinsic parameters. As such, in our opinion, it is reliable to conduct parameter estimation for highly precessing binary black holes up to mass ratios of $4:1$ and SNRs of 40 with \SEOB{}.
Up to mass ratios of $2:1$, \TPHM{} and \XPHM{} also provide robust recovery though not necessarily as
reliable as \SEOB{} on average. Nonetheless, as can be seen for several marginalized posterior distributions in Fig.~\ref{fig:three_model_posteriors}, these two models outperform \SEOB{} on occasion. This reminds us of the necessity of multi-model inference. However, the general trends of our results show that \SEOB{}
is in general more reliable and thus should be given more weight in multi-model PE when appropriate.
We explore this in detail in Sec.~\ref{Sec:NR_informed_PE}.

Let us conclude this section with the interesting observation that the $\{M,q,\chieff,\chip\}$ posteriors 
from the $Q=2$ subset are wider than their corresponding $Q=1,4$-subset posteriors. 
This holds for each model as can be gathered from Table~\ref{tab:width_ratio}. This may be because the likelihood gets more sharply peaked with decreasing $q$ at a fixed SNR.
Indeed, explicitly comparing the width of the $q=1/2$ posteriors with those of the $q=1/4,1$
posteriors, we find the former to be $30\%$ to twice larger on average 
than the latter two.
This can also be gleaned from Refs.~\cite{Pratten:2020ceb, Estelles:2021gvs} though it is less clear-cut as the total SNRs are 
different, as would be the precession SNRs.
Since $\chieff$ and $\chip$ are both functions of $q$, this behaviour is also present in
their posteriors as seen in Table~\ref{tab:width_ratio} and Fig.~\ref{fig:three_model_posteriors}. 
This is even manifest in the $\Mtot$ posteriors due to the fact that
what is directly inferred is the chirp mass $M_c$, which gives us the total mass via another mass-ratio-dependent relation
$\Mtot = [q/(1+q)]^{-3/5}M_c$. Therefore, the resulting $\Mtot$ posteriors for the $Q=2$ subset are wider than their $Q=1,4$ counterparts as corroborated once again by
Table~\ref{tab:width_ratio} and Fig.~\ref{fig:three_model_posteriors}.

\subsection{Cases with mass ratio $8:1$}
\label{sec:Q8_injections}

Due to the abundance of the publicly available \SXS{} simulations with $Q\le 4$, 
performance studies of waveform models usually stop at this mass ratio.
However, we wish to assess the performance of the models for BBHs with more asymmetric masses.
To this end, we select five simulations from the BAM catalog.
These are \BAM{67}, \BAM{69}, \BAM{73}, \BAM{77} and \BAM{78}. Their relevant intrinsic
parameters are provided in Table~\ref{tab:BAM_sim_params} which makes it clear that these are all single-spin simulations.
Note that we chose this particular subset of simulations so as to slowly increase $\chip$ from $0.346$ for \BAM{67} to $\chip = 0.801$  for \BAM{78}
while still keeping the total SNR at 40 with the precession SNR $\rho_{\mathrm{p}}$ at 10.

We begin with Fig.~\ref{fig:three_model_posteriors_BAM} which displays the three-model recovery posteriors
for the set of parameters $\{\Mtot, q,\chieff,\chip\}$ resulting from the injection of the $Q=8$ BAM simulations listed in Table~\ref{tab:BAM_sim_params}.
The color coding is the same as in Fig.~\ref{fig:three_model_posteriors}.
For the recovery of the binary total mass, \SEOB{} and \TPHM{} exhibit alternating performances in the sense
that if one model recovers $\Mtot$ within $\pm 1\sigma$, the other does so within $\pm(2$-$3)\sigma$.
Regardless, the combined posteriors between the two models do recover $\Mtot$ reliably.
These statements also hold true for the \SEOB-\TPHM{} recovery of the mass ratio $q$ as can be seen
in the second column of Fig.~\ref{fig:three_model_posteriors_BAM}.

Despite similarities in their recovery of $\{\Mtot,q\}$, \SEOB{} outperforms \TPHM{} in inferring $\chieff$.
This is mostly because of the aforementioned drop in \TPHM's in performance for when spin magnitude(s)
are somewhat large and tilt angle(s) very close to $\pi/2$. 
Indeed, the two simulations, \BAM{73} and \BAM{78}, for which \TPHM{} overestimates $\chieff$ by
more than $3\sigma$ have $a_1\ge 0.6$ and $\theta_1\approx \pi/2$ resulting in $\vert\chieff\vert \lesssim 10^{-2}$ while $\chip \gtrsim 0.6$.
We had already highlighted this issue with \TPHM{} for the $Q=4$ cases in Sec.~\ref{sec:Q4_subset}
and it seems to have become even more severe for the $Q=8$ cases.
\SEOB{} and \TPHM{} again behave similarly in their recovery of $\chip$ with the two models
unable to recover $\chip$ reliably either individually or combined together\footnote{Here, by ``reliably'' we mean that the posteriors should recover $\chip$ within $\pm 2\sigma$ for all five injections.}.
This necessitates further model calibration in this regime of the intrinsic parameter space.

\XPHM's recovery of $\{\Mtot,q\}$ is robust for the two cases with $a_1=0.4$.
It seems to also recover $\chieff$ within $\pm 2\sigma$ for these two cases,
but once the spin magnitude exceeds $0.4$, the model's reliability drops significantly.
However, it recovers $\chip$ well for the two cases that are most challenging for \SEOB{}
and \TPHM:  \BAM{73} and \BAM{78} with $a_1\ge 0.6$ and $\theta_1\approx \pi/2$.
We should caution that this is not a conclusive finding since it is based on two cases.

Interestingly, the combined \SEOB-\TPHM{} posteriors reliably recover $\{\Mtot,q,\chieff\}$
for all five cases and recover $\chip$ only for three of the five simulations.
Therefore, at least for strongly single-spin precessing BBHs with high mass asymmetry,
we advise use of at least \SEOB{} and \TPHM{} in the parameter estimation.
More injection-recovery studies should be conducted at this mass ratio for establishing
efficient PE strategies for future events.

In summary, waveform model recovery performance deteriorates further as we go from mass ratio $4:1$
to mass ratio $8:1$ systems. \SEOB{} still provides marginally better recoveries than \TPHM{}
while \XPHM{} mostly produces posteriors that are biased by more than two standard deviations. 
We hypothesize that the reason for this trend is due to higher order multipoles in the GW emission which become more prominent as the mass ratio of the binary increases~\cite{Mills:2020thr}. Since \XPHM{} does not include calibration to numerical relativity simulations for the higher multipoles in precessing systems, it is possible that errors in these high-multipole contributions could cause biases in our parameter estimation analyses.
As the amplitude of each higher multipole becomes more significant, this will become more of an issue. However, we note that \SEOB{} and \TPHM{} are also not calibrated to numerical relativity simulations for the higher multipoles in precessing systems yet both models perform better than \XPHM{}. Nevertheless,
what is clearly evident is that no single model can be said to be reliable in this regime.
Now that many more precessing $Q=8$ simulations have become available \cite{Scheel:2025jct}, a study similar to our $Q=4$ subset
should be conducted on these models.

\section{Results II: Inspiral-Merger/Ringdown Consistency Tests of the Models}
\label{Sec:IMR_tests}

We begin with a qualitative description of the Inspiral-Merger/Ringdown (IMR) consistency test. We highlight how estimates for the properties of the remnant black hole can be made from independent measurements from the inspiral and ringdown portions of the signal and therefore checked for consistency.

The early inspiral portion of the GW signal is well described as a post-Newtonian series with the chirp mass $\mathcal{M}_c$ contributing \new{to the GW phasing} at the leading (Newtonian) order and
the symmetric mass ratio $\eta$ entering \new{at $1/2$ PN order higher}.
The spins \new{first enter at the relative 1.5PN order} via their components parallel to the angular momentum: $\chi_i=a_i\cos\theta_i$, for $i=1,2$ \cite{Poisson:1995ef, Blanchet:2013haa}. Based solely on the amplitude, phase and frequency evolution of the inspiral portion of the GW signal, estimates for the binary parameters can be obtained~\cite[see e.g.][]{Fairhurst:2023idl}. These parameters can be mapped to estimates for the properties of the remnant black hole by evaluating numerical-relativity fits~\cite{Jimenez-Forteza:2016oae, Healy:2016lce, Hofmann:2016yih}, as has been done previously~\cite{Ghosh:2016qgn,Ghosh:2017gfp,Breschi:2019wki}.

On the other hand, the ringdown portion of the GW signal is well described by black hole perturbation theory \cite{Regge:1957td, Vishveshwara:1970zz, Zerilli:1971wd, Press:1971wr, Davis:1971gg, Chandrasekhar:1975zza, Detweiler:1977gy, Kokkotas:1999bd, Berti:2025hly}. Here, a perturbed Kerr black hole radiates GWs via a superposition of
exponentially damped quasinormal modes (QNMs), where the QNM frequencies and damping times are functions
of only the mass and spin of the unperturbed final Kerr black hole (see Refs.~\cite{Kokkotas:1999bd, Berti:2009kk, Franchini:2023eda} for reviews). 
Based solely on a measurement of the frequency and the decay time of the fundamental QNM, estimates for the binary parameters can be obtained~\cite{Echeverria:1989hg, Finn:1992wt},

We therefore have two independent methods for obtaining the final mass and spin of the remnant black hole.
In the case of zero noise and assuming GR to be the correct theory of gravity, 
these measurements should agree modulo modelling systematics.
To quantify this agreement, we introduce the dimensionless quantities \cite{Ghosh:2017gfp} 
\be
\Delta M_f/\bar{M}_f,\quad  \Delta a_f/\bar{a}_f , \label{eq:Delta_M_delta_a}
\ee
where
an overbar denotes the average between the inspiral (\insp) and the merger-ringdown (MR) values, viz.,
\begin{align}
\bar{M}_f &:= \left(M_f^\insp + M_f^\text{MR}\right)/{2}, \label{eq:M_f_bar}\\
\bar{a}_f &:= \left(a_f^\insp + a_f^\text{MR}\right)/{2} \label{eq:a_f_bar}
 \end{align}
and
\begin{align}
 \Delta M_f &:= M_f^\insp-M_f^\text{MR}, \label{eq:Delta_M_f}\\
\Delta a_f &:= a_f^\insp-a_f^\text{MR} \label{eq:Delta_a_f}.
 \end{align}
If there are no systematic biases due to mismodelling then we expect the posteriors for 
$\Delta M_f/\bar{M}_f$ and $\Delta a_f/\bar{a}_f$
to be centered around zero \cite{TheLIGOScientific:2016src}.
In this case, the inspiral values from post-Newtonian theory are said to be consistent with those from the ringdown based on black hole perturbation theory.

Given that waveform models employ fitting formulas calibrated to NR simulations to obtain $M_f, a_f$ as functions of the initial masses and spins, 
the IMR consistency check is a good way to expose systematic biases.
Such fits were already used to estimate $M_f, a_f$ for GW150914 \cite{Abbott:2016blz}.
Spin precession only complicates the matter as, e.g., the waveform models extend the 
post-inspiral Euler angles (used in the frame rotation to go to the final $a_f$-frame) in different ways 
\cite{Pratten:2020ceb, Estelles:2020twz, Hamilton:2021pkf, Gamba:2021ydi, Ramos-Buades:2023ehm, Thompson:2023ase, Colleoni:2024knd}.
An especially alarming scenario is systematic biases mimicking false beyond-GR signatures in parameter estimation \cite{Gupta:2024gun, Dhani:2024jja}\footnote{Alternatively, an eccentric BBH signal analyzed with quasicircular templates can also yield a false beyond-GR signature \cite{Bhat:2022amc}.}.
Therefore, it is of crucial importance to systematically survey the performance of different waveform models under the IMR consistency test.

This test has become a standard check for the LVK collaboration
\cite{LIGOScientific:2019fpa, Breschi:2019wki, LIGOScientific:2020tif, LIGOScientific:2021sio}.
The details of the implementation within a Bayesian framework can be found in Ref.~\cite{Ghosh:2016qgn} (also see Refs.~\cite{TheLIGOScientific:2016src, Ghosh:2017gfp, Carson:2020rea, Krishnendu:2021fga}), so we provide a brief introduction here.
First, we must select a cutoff frequency, $\bar{f}$, to separate the inspiral regime from the MR part of the signal. One option is to set
\be
\bar{f}= f_\text{ISCO}^\text{Sch}= (\pi)^{-1} 6^{-3/2}\left(G\Mtot/c^3\right)^{-1}\approx (60\Msun/\Mtot)\, 73\,\text{Hz} \label{eq:Sch_ISCO_freq},
\ee
i.e., the GW frequency corresponding to the innermost stable
circular orbit (ISCO) in Schwarzschild spacetime.
We have checked that this is less than the frequency of
the minimum energy circular orbit (MECO) \cite{Pratten:2020fqn} for all our injections using the
\texttt{lal.SimIMRPhenomXfMECO} function.
Another choice is to set
\be
\bar{f}=f_\text{ISCO}^\text{Kerr}= \f{c^3}{\pi G}\f{\sqrt{M}}{r_\text{ISCO}^{3/2}+a\sqrt{M}} \label{eq:Kerr_ISCO_freq},
\ee
where $r_\text{ISCO}$ is obtained from the solution to \cite{Bardeen:1972fi}
\be
 r(r-6M)+8a\sqrt{Mr}-3a^2 = 0 .\label{eq:ISCO_cubic}
\ee
This second choice corresponds to the ISCO frequency of an equatorial, prograde timelike orbit around a Kerr black hole of mass $M$ and
spin $a=\chi M$ with $\chi \in [0,1)$~\cite{Ghosh:2016qgn}.

%==============================================================
%    FIG: XPHM Pulsar Fig horizontal
%==============================================================
\begin{figure*}[t!]
    \centering
    \hspace*{-3em}\includegraphics[width=1.1\textwidth]{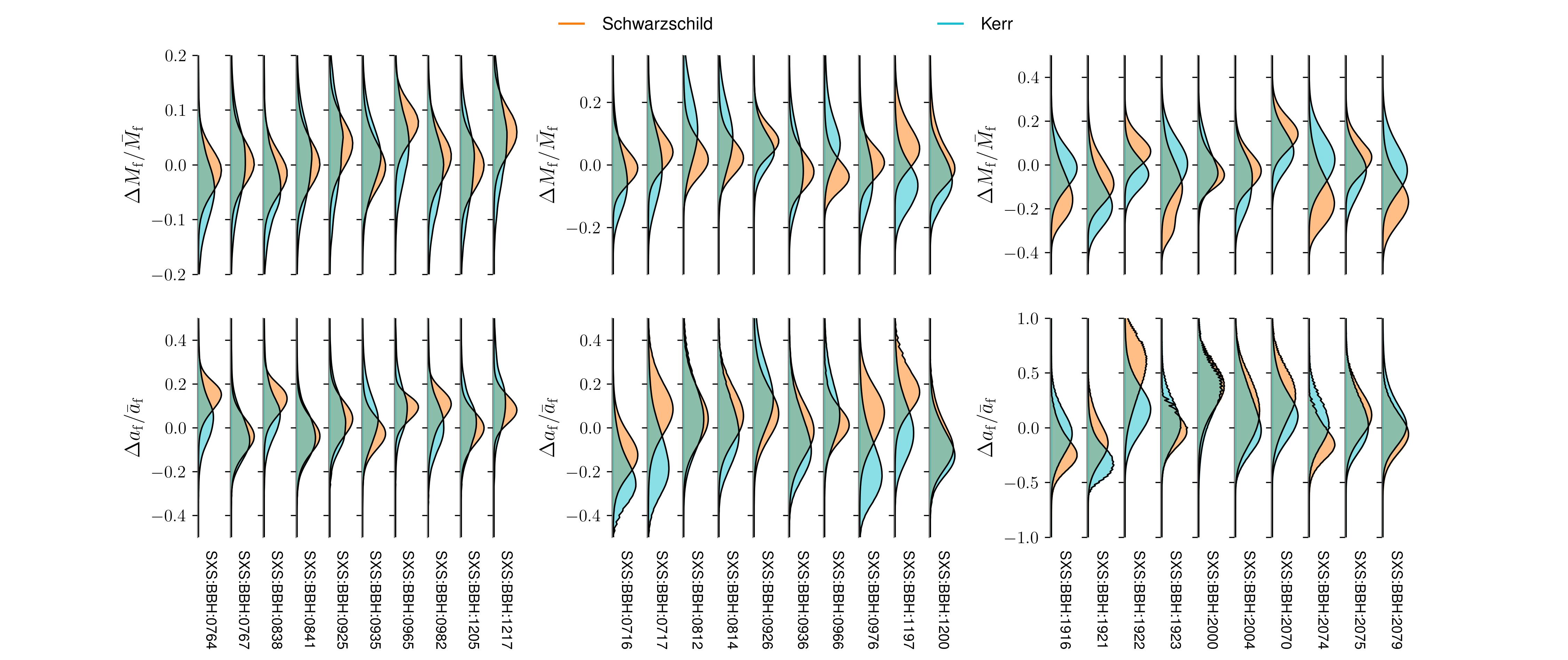}
\caption{The posteriors for $\Delta M_f/\bar{M}_f$ (top row) and $\Delta a_f/\bar{a}_f$ (bottom row)
from the IMR consistency test performed with \XPHM{} for the 30 $Q\le 4$ numerical relativity
simulations whose parameters are given in Tables~\ref{tab:SXS_sim_params} and \ref{tab:SXS_extrinsic_params}. 
See Eqs.~(\ref{eq:M_f_bar}-\ref{eq:Delta_a_f}) for the relevant definitions.
Each row is further grouped into three clusters of 10 subfigures each separated by the mass ratio
of the binary black hole simulations with $Q=1,2,4$ clusters from left to right.
Note that the vertical scale is different in each cluster. In each plot, we show the IMR results obtained with two different choices for the cutoff frequency, see Eqs.~(\ref{eq:Sch_ISCO_freq},\ref{eq:Kerr_ISCO_freq}). In orange (cyan) we show the results obtained with a cutoff frequency equal to the Schwarzschild (Kerr) ISCO frequency.}
\label{fig:XPHM_IMR_30sims_horizontal}
\end{figure*}

Based on GW signals with only the dominant quadrupolar contribution Ref.~\cite{Ghosh:2016qgn} argues that
the specified cutoff frequency should not have a significant effect on the IMR consistency test, as long as a reasonable selection
is made.
We test their assertion here for waveform models that include precession effects and
higher multipoles by re-conducting the same IMR consistency tests with both of these frequency
cutoffs.
For a non-spinning BBH with $q=1,a_1=a_2=0$, the resulting black hole's final spin
is approximately $2M/3$ \cite{Tichy:2008du}. Thus, for $M=60\Msun, a=2M/3$, Eq.~\eqref{eq:Kerr_ISCO_freq} gives us approximately $150\,$Hz, roughly twice the value
from Eq.~\eqref{eq:Sch_ISCO_freq}.
We list both cutoff frequencies in Table~\ref{tab:GR_quantiles} in App.~\ref{Sec:App_A}, where one sees that
the Kerr frequency is, in general, $2$ to $2.5$ times larger than the Schwarzschild one.

We note that a $60\Msun$, $a=2M/3$ Kerr black hole has
its $\ell=2$ fundamental ($n=0$) QNM frequencies ranging from $\sim 150\,$Hz to $\sim 250\,$Hz \cite{Berti:2009kk}, the former being the QNM frequency of the $(2,-2)$ multipole, and the latter $(2,2)$. So although the Kerr ISCO may be an appropriate cutoff for IMR consistency
tests involving only the $(2,2)$ mode, and indeed is commonly used by the LVK~\cite{LIGOScientific:2021sio}, this cutoff may be too high when other higher order multipoles are considered. Be that as it may, we explore how the results of the
IMR consistency test change depending on the choice of cutoff.

For the inspiral PE analyses, the likelihood is integrated from 23\,Hz to $\bar{f}$ (the lower bound is consistent with the full IMR analyses, see Sec.~\ref{Sec:injection_primer} for details).
For the merger-ringdown analyses, we integrate the likelihood from $\bar{f}$ to 2048\,Hz.
The final mass and spin from the inspiral part are obtained from the posteriors
for the binary masses and spins via waveform-specific fits to NR simulations \cite{Jimenez-Forteza:2016oae, Healy:2016lce, Hofmann:2016yih}.
$M_f,a_f$ are also derived from the inferred fundamental QNM of the MR injection analyses.
As in Sec.~\ref{Sec:recovery_3models}, our Bayesian inference analyses are performed using the {\texttt{bilby} and {\texttt{bilby\_pipe} packages
with the {\texttt{dynesty} sampler (using consistent settings as before).
Given that we are using consistent priors as in Sec.~\ref{Sec:recovery_3models}, this leads to non-uniform effective priors for $\Delta M_f/\bar{M}_f$ and $\Delta a_f/\bar{a}_f$. Ref.~\cite{LIGOScientific:2020tif} re-weighted the inspiral and MR analyses to uniform priors for the deviation parameters, arguing that this more clearly conveys the information gained from the data. Consistent with Ref.~\cite{LIGOScientific:2019fpa}, we do not perform any re-weighting in this work. We finally use the
{\texttt{pesummary}} package~\cite{Hoy:2020vys} to calculate posteriors for
$\Delta M_f/\bar{M}_f$ and $\Delta a_f/\bar{a}_f$, via the {\texttt{summarytgr}} executable. 

\subsection{Cases with $Q\le 4$}
\label{sec:IMR_test_Qle4}
For expediency, we perform the IMR consistency test of all 30 NR injections with \XPHM{}.
We present results using other waveform models for selected cases owing to computational restrictions.
The 1D marginalized posteriors for $\Delta M_f/\bar{M}_f$ and $\Delta a_f/\bar{a}_f$
from the \XPHM{} recovery of the 30 $Q\le 4$ injections using both cutoff frequencies
are displayed in the top and bottom rows of Fig.~\ref{fig:XPHM_IMR_30sims_horizontal}.
Each row is further divided into three sections containing 10 plots separated by the mass ratio
with the $Q=1\ (Q=4)$ subset on the left (right).

To quantify the performance of \XPHM{} in the IMR consistency test, we additionally introduce a new metric for the posterior recovery in the context
of the IMR consistency test: $\mathcal{D}^{\mathrm{2D}}$.} This is the probability that gives the fraction of the 2D posterior probability distribution enclosed by the isoprobability contour that passes through $0$ (see, e.g., Ref.~\cite{LIGOScientific:2016lio} for details). Under this metric, smaller probabilities indicate better consistency with GR.
For example, a 2D normal distribution that captures 0 within two standard deviations along a single dimension (with the cross-section along second dimension remaining centered around 0) yields $\mathcal{D}^\text{2D} \leq 88\%$; this increases to $\mathcal{D}^\text{2D} \leq 99\%$ if the 2D distribution captures 0 within two standard deviations in both dimensions.
We list the values we obtain for $\mathcal{D}^\text{2D}$ for all 30 cases using both
cutoff frequencies in Table~\ref{tab:GR_quantiles}.

Note that
GR deviations
are seen more frequently when studying the marginalized 1D posterior distributions
than the 2D posteriors. Consider, e.g., \sxsSim{0965} whose 1D posteriors in  Fig.~\ref{fig:XPHM_IMR_30sims_horizontal} for both $\Delta M_f/\bar{M}_f$ and
$\Delta a_f/\bar{a}_f$ point to GR violations using the Schwarzschild ISCO cutoff.
However, the corresponding value for $\mathcal{D}^\text{2D}$ in Table~\ref{tab:GR_quantiles} is $85\%$ indicating, somewhat marginally, that the final
mass and spin are consistent with GR.
Since a true-positive GR violation is a rather unexpected result, we opt for the
most conservative metric: we declare the IMR consistency test (IMRCT) a failure
only when $\mathcal{D}^\text{2D} > 88\%$ \emph{and} when both 1D posteriors ``miss''
the GR value, 0, by more than two standard deviations.
This is still a frequency dependent claim as we explain below.

The first observation from Fig.~\ref{fig:XPHM_IMR_30sims_horizontal} is that, in general,
the posteriors obtained using two different cutoffs are not consistent.
In other words, the results of the IMRCT \emph{depend} on the choice
of the cutoff frequency, at least for signals composed of multiple harmonics emitted by strongly precessing BBHs. The 1D posteriors obtained using the cutoff based on the Kerr black hole are,
in general, broader than the posteriors obtained using the Schwarzschild cutoff. This is reasonable since there is less information/SNR in the post-inspiral phase of the signal when employing the larger Kerr cutoff, and therefore wider posterior distributions are expected.
As a result, the Kerr ISCO results tend to be more consistent with GR as confirmed by the values
of $\mathcal{D}^\text{2D}$ in Table~\ref{tab:GR_quantiles}.
However, for the $Q=2$ subset of BBHs, we report that in general the Schwarzschild ISCO results are more consistent with GR.

The discussion above raises the question of whether a single clear case of synthetic GR violation can be found. For example, using the Kerr ISCO cutoff
and the 1D cyan posteriors of Fig.~\ref{fig:XPHM_IMR_30sims_horizontal},
we could declare a GR violation for \sxsSim{1921}. However, we see that the results
using the Schwarzschild cutoff contradict this.
Returning to Table~\ref{tab:GR_quantiles}, we observe that there is not a single
occurrence of both $\mathcal{D}^\text{2D}_\text{Sch}$ and $\mathcal{D}^\text{2D}_\text{Kerr}$ exceeding $88\%$ for the same case.
If we relax this bound somewhat, we can place \sxsSim{0838, 2000} in a list
of cases of interest.
Looking at the 1D posteriors in Fig.~\ref{fig:XPHM_IMR_30sims_horizontal}, we can expand
this list to also include \sxsSim{0926, 0965, 1197, 1217, 1922, 2070} giving
us eight special cases to explore in greater detail. Given that in some cases we observed large differences in the obtained posterior distributions for different waveform models in Sec.~\ref{Sec:recovery_3models}, two natural questions are
\begin{inparaenum}[(a)]
\item how much the differences between the Schwarzschild and Kerr cutoff frequencies depend on the choice of a particular waveform model, and
\item whether or not model systematics can impact GR violation statements.
\end{inparaenum}
To this end, we re-ran the IMRCT with \TPHM{} and \SEOB{} for the above-list of eight cases.

With regards to the first question, we obtained consistent conclusions for \TPHM{} and \SEOB{} as with \XPHM{}, {i.e.}, the Schwarzschild ISCO posteriors are narrower than the Kerr results, and the Kerr results yield posteriors that are more consistent with GR
as shown in Fig.~\ref{fig:IMR_comparison_SEOB} in App.~\ref{Sec:App_A}.
We next focus on the second question. Rather than presenting results from two different cutoff frequencies, which show broadly the same features among the waveform models considered, we solely focus on the Schwarzschild ISCO results since they show the worst case scenario.
In Fig.~\ref{fig:IMR_comparison}, we show a comparison of the performance of these three models in the IMRCT.
We see significant differences between the models,
especially in the $\Delta a_f/\bar{a}_f$ posteriors. It is worth noting that when this analysis is repeated with the Kerr cutoff, we observe more consistent results owing to the wider posterior distributions.
It is clear from Fig.~\ref{fig:IMR_comparison} that \SEOB{} and \TPHM{} perform better in the IMRCT
than \XPHM. Moreover, \TPHM{} recovers both $\Delta M_f/\bar{M}_f$ and $\Delta a_f/\bar{a}_f$
within $\pm 2\sigma$ every time. $\Delta a_f/\bar{a}_f$ consistency is especially robust as evident from the last column of the figure.
\SEOB{} can also be said to mostly pass this test, but it seems to produce less consistent final spins for
more mass symmetric, i.e., $Q\le 2$ binaries than \TPHM, while its posteriors of $\Delta a_f/\bar{a}_f$
for the three $Q=4$, shown in Fig.~\ref{fig:IMR_comparison}, are very consistent.
Perhaps this is indicative of the model's performance toward the single-spin limit which we explore more
in Sec.~\ref{sec:IMR_test_Q8}.
\begin{figure}[t!]
    \centering
    \includegraphics[width=0.49\textwidth]{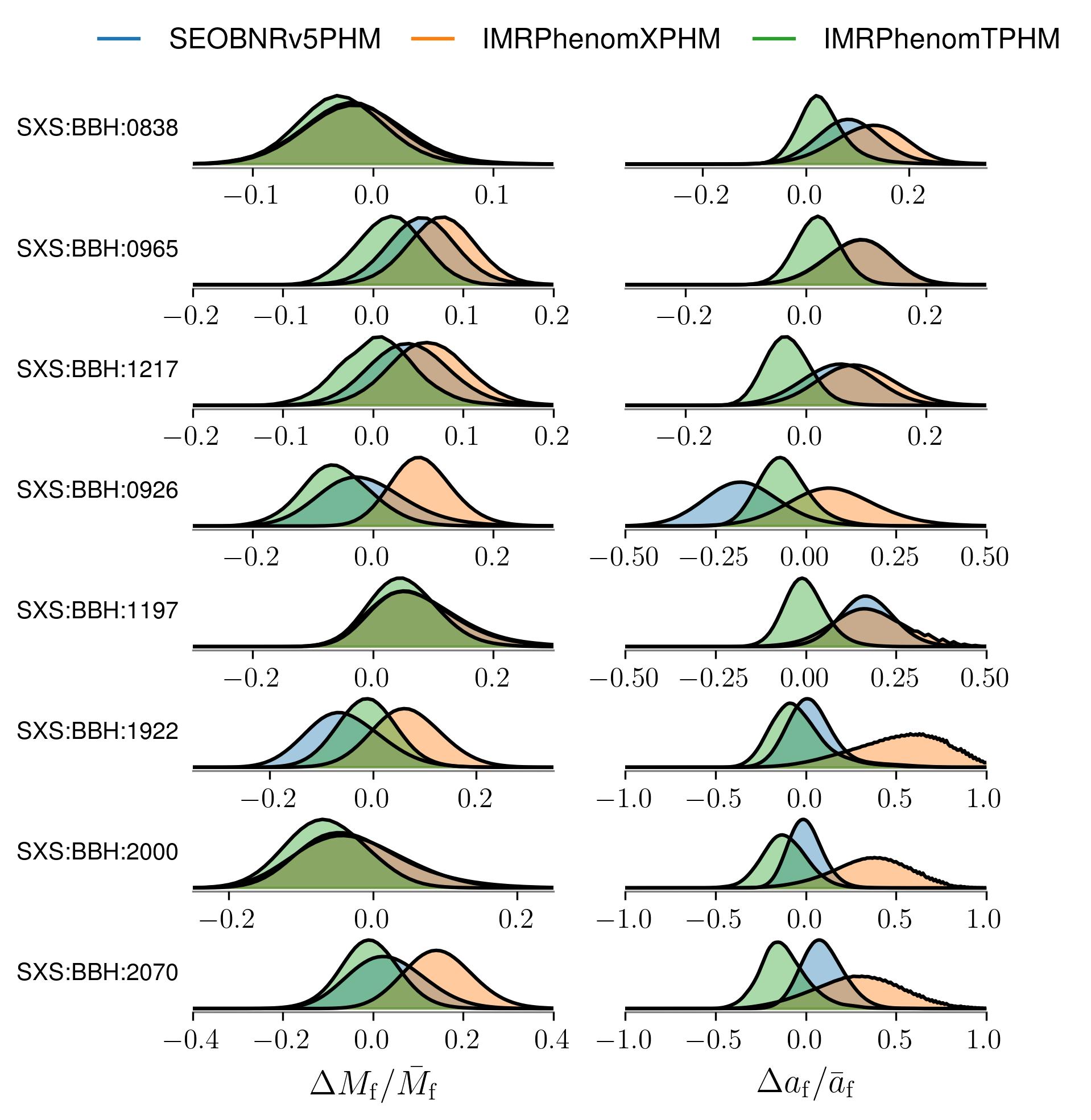}
\caption{Model performance under the IMR consistency test for the eight special cases
mentioned in Sec.~\ref{sec:IMR_test_Qle4} using the Schwarzschild ISCO as the frequency cutoff [Eq.~\eqref{eq:Sch_ISCO_freq}].
}
\label{fig:IMR_comparison}
\end{figure}

We additionally checked the performance of the recently developed frequency-domain model
\modelname{IMRPhenomXO4a} (\XOFourA) \cite{Thompson:2023ase} for these eight IMRCTs. We present the results in Fig.~\ref{fig:IMR_comparison_XPHM_XO4a} in App.~\ref{Sec:App_A}} where we compare this
model's performance with that of \XPHM.
\XOFourA{} seems to bring some improvement with the most important differences being in the posteriors for $\Delta a_f/\bar{a}_f$ for the three $Q=4$ cases: \sxsSim{1922, 2000, 2070}.
In particular, \XOFourA's posteriors are much narrower and peak much closer to $\Delta a_f=0$ than for \XPHM. However, the former are still biased by $\gtrsim 2\sigma$.

As for the drop in \XPHM's performance for these eight cases, and especially the
three $Q=4$ cases, this
may be due to the mismodelling of QNM frequencies in the co-precessing frame which is exacerbated for BBHs with $Q\gtrsim 4,a_1 \gtrsim 0.7$ and $\theta_1 \gtrsim \pi/2$,
and is worst for $\theta_1\approx 150^\circ$ in the single-spin case \cite{Foo:2024exr}.
This is consistent with the findings of Ref.~\cite{Hamilton:2023znn} where it is shown that
for single-spin systems, $\theta_1\approx 150^\circ$ corresponds roughly to $\cos\langle\beta\rangle \approx 0$
where $\beta = \cos^{-1}(\hat{\mathbf{J}}\cdot \hat{\mathbf{L}})$ and the angular brackets represent
its average in the frequency domain between $f(t=t_c)$ and $f(t=t_c+\Delta t)$ with $t_c$ denoting the coalescence
(merger) time and $\Delta t \in [40,90]\Mtot$. Essentially, when $\cos\beta=0$, the shift away from the true QNM ringdown frequency is
maximized \cite{Foo:2024exr}.
\new{Further support for this hypothesis is the fact that both \SEOB{} and \XOFourA{} have improved merger-ringdown attachments based on the work of Ref.~\cite{Hamilton:2023znn}. This might explain why they produce $\Delta a_f/\bar{a}_f$ posteriors more consistent with GR than \TPHM{} for the aforementioned three $Q=4$ cases.
However, \TPHM{} has a similar ringdown prescription to \XPHM, but performs better for the $Q=4$ BBHs under the IMRCT. So the missing QNM corrections offer a partial explaination for the \XPHM{} results.}
We will return to this discussion in the next section,
where we consider systems even closer to the single-spin limit ($Q=8$).

Focusing on \XPHM{} with the Schwarzschild ISCO cutoff,
we performed a few more checks motivated by the results in
Ref.~\cite{Narayan:2023vhm},
where they observe GR violations even when injecting and recovering with the same waveform model.
They argue that the presence of higher-order multipoles may be the cause as there have not been extensive tests on the IMRCTs for waveform models including higher-order multipoles. To investigate whether this may be the reason why we observe inconsistent recoveries for $M_f$ and $a_f$, we re-perform the IMRCT, but now
{inject \XPHM{} waveforms that have} the same parameters as the {previously-injected \SXS{} waveforms}. By doing so, we remove the possibility of model systematics biasing results. {In this fashion,} we re-analyze these eight aforementioned \SXS{} simulations.
As can be seen in Fig.~\ref{fig:imr_comparison_nr_vs_xphm} in App.~\ref{Sec:App_A}, we find that in general \XPHM{} performs much better in the IMRCT for simulations that are produced using \XPHM{} {injections} compared with NR. This implies that the $M_f$ and $a_f$ inconsistencies that we observe in Fig.~\ref{fig:XPHM_IMR_30sims_horizontal} are indeed due to model systematics, and hence a consequence of \XPHM{} not faithfully representing NR for strongly precessing GW signals.

We recommend that these eight simulations be used as the IMR consistency test set for future precessing waveform models, keeping in mind that the results are sensitive
to the choice of the cutoff frequency.

\subsection{Cases with mass ratio $8:1$}\label{sec:IMR_test_Q8}

As in the last section, we begin with results obtained using \XPHM.
These are shown in Fig.~\ref{fig:IMR_comparison_XPHM_Q8} for the five $Q=8$ injections of Table~\ref{tab:BAM_sim_params}.
We see from the figure that the posteriors are quite distinguishable from
each other, i.e., the IMRCT results continue to be sensitive to the cutoff
choice as $q$ decreases.

In Fig.~\ref{fig:IMR_comparison_TPHM_Q8}, we show the results of the IMR consistency test
on the three waveform models using the Schwarzschild ISCO cutoff.
As was the case with the $Q\le 4$ injections, \TPHM{} produces the most consistent posteriors for both $\Delta M_f/\bar{M}_f$ and $\Delta a_f/\bar{a}_f$ in four out of the five cases. \SEOB{} performs similarly for spin consistency,
but yields $\Delta M_f/\bar{M}_f$ posteriors for \BAM{67} that are inconsistent at $3\sigma$. \XPHM{} produces very broad, mostly uninformative, posteriors for $\Delta a_f/\bar{a}_f$, but is, in principle, consistent with $\Delta a_f=0$ to $2\sigma$ for all five simulations.
On the other hand, it consistently points to $\Delta M_f <0$ at $1\sigma$ to $3 \sigma$.

\begin{figure}[t!]
    \centering
    \includegraphics[width=0.49\textwidth]{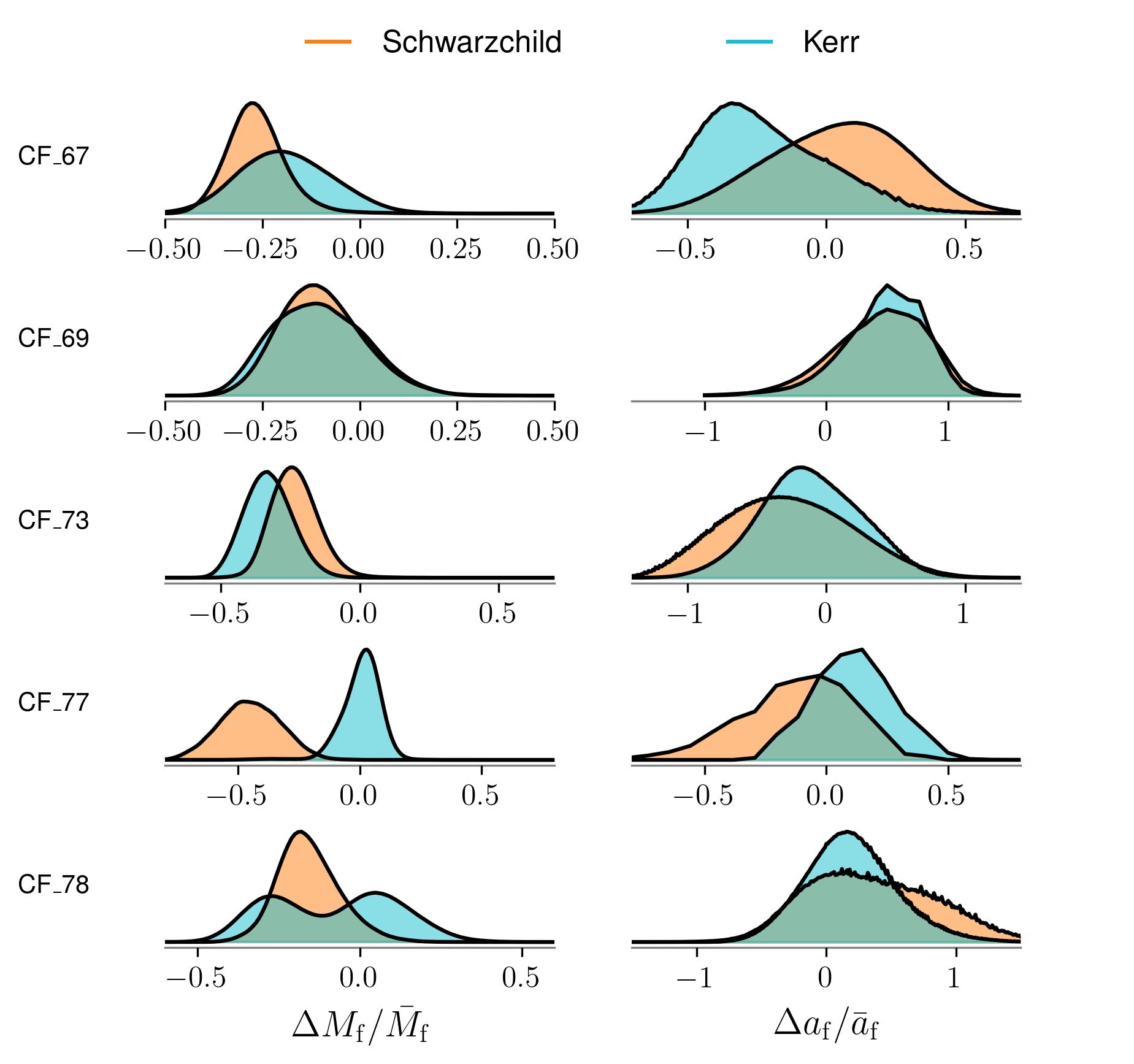}
\caption{Results of the IMR consistency test of Sec.~\ref{Sec:IMR_tests} with \XPHM{} for our $Q=8$ binary black hole
injections whose parameters are given in Table~\ref{tab:BAM_sim_params}. In each plot, we show the IMR results obtained with two different choices for the cutoff frequency, as in Fig.~\ref{fig:XPHM_IMR_30sims_horizontal}.}
\label{fig:IMR_comparison_XPHM_Q8}
\end{figure}

Previous studies have shown that $Q\gtrsim 4,a_1 \gtrsim 0.7,\theta_1 \approx 150^\circ$ is the most challenging for \XPHM{} in the IMRCT \cite{Hamilton:2023znn, Foo:2024exr}. In our sample set, the simulation \BAM{69} with $Q=8, a_1=0.4$ and $ \theta_1\simeq 120^\circ$ is has parameters closest to the region. Consistent with previous works, we observe the worst [2D] recovery using both cutoffs for this simulation as shown in Fig.~\ref{fig:IMR_comparison_XPHM_Q8}.
On the other hand, the \XPHM{} posteriors recovered for \BAM{77} are mostly more consistent with GR despite this simulation having $a_1=0.8$, possibly because $\theta_1\simeq 60^\circ$.

Another interesting finding is the inability of the models to pass the $\Delta M_f$ consistency
check for \BAM{73} when using the Schwarzschild ISCO cutoff.
\SEOB{} and \TPHM{} also fail the test for $\Delta a_f$ and \XPHM{}
produces a posterior for $\Delta a_f/\bar{a}_f$ that ranges from $-1$ to $1$.
This simulation has $a_1=0.6$ and $\theta_1\simeq 90^\circ$, so it has rather strong precession, but not as much as \BAM{78} with $a_1=0.8,\theta_1\simeq 90^\circ$,
which produces much more consistent results.
We performed several additional checks/runs to understand this behaviour:
\begin{inparaenum}[(i)]
 \item
 We visually compared the 90\%-credible time-domain waveforms generated by the three models for \BAM{73} to those for \BAM{78} and found no smoking-gun signatures explaining the issue.
 \item
 We re-injected the BBH with intrinsic parameters of \BAM{73} (\BAM{78}) and extrinsic parameters of \BAM{78} (\BAM{73}) and recovered posteriors consistent with $\Delta M_f=\Delta a_f=0$
 for all three models similar to what is shown for \BAM{78} in Fig.~\ref{fig:IMR_comparison_TPHM_Q8} (again using the Schwarzschild ISCO cutoff).
 \item
 We computed the precession SNR, $\rho_\text{p}$, for these two new injections, as well as the original injections of \BAM{73} and \BAM{78}. The troublesome case yields the lowest value with $\rho_\text{p}=9.6$ with the original \BAM{78} injection giving $\rho_\text{p}=9.9$ and the two new injections mentioned
 in (ii) yielding $11.0$ and $10.7$, respectively.
 \item
 We finally looked at how the injected SNR is distributed across the LIGO L1-H1 network.
 For \BAM{73}, we have SNR$_\text{H1}=11$, SNR$_\text{L1}=38$; for \BAM{78},
 SNR$_\text{H1}=31$, SNR$_\text{L1}=25$. However, for the re-injections of (ii), we
 have $31, 25$ and $10, 38$, respectively, consistent with the original cases. 
 \end{inparaenum}
In short, we have not been able to determine why the models pass the IMRCT for \BAM{78},
but not for \BAM{73}.
\begin{figure}[t!]
    \centering
    \includegraphics[width=0.49\textwidth]{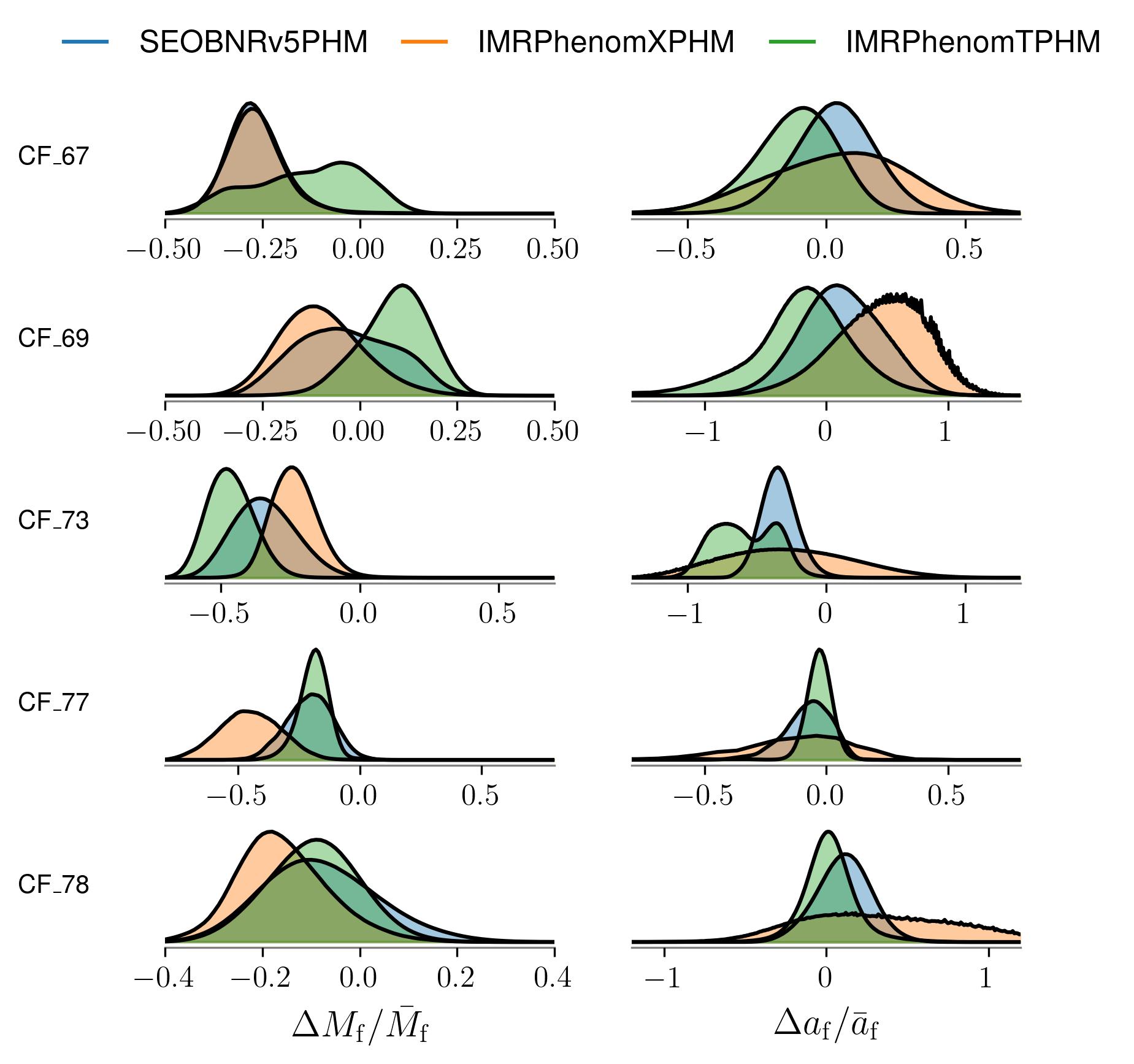}
\caption{Model performance under the IMR consistency test for our $Q=8$ binary black hole
injections whose parameters are given in Table~\ref{tab:BAM_sim_params} and discussed in Sec.~\ref{Sec:IMR_tests}. For these cases we consider a cutoff frequency based on the Schwarzschild ISCO.}
\label{fig:IMR_comparison_TPHM_Q8}
\end{figure}
In summary, for the strongly double-spin precessing BBHs considered here,
the results of the IMR consistency test depend on the choice of the cutoff frequency. The posteriors obtained using the cutoff based on the Kerr ISCO are generally
more consistent with GR, albeit broader. The results obtained using the cutoff based
on the Schwarzschild ISCO can be taken to be more informative though can sometimes point to false deviations of GR.

Focusing on the Schwarzschild cutoff, we see that
\TPHM{} passes the IMR consistency test convincingly for mass ratios up to $Q=4$. \SEOB{} mostly produces consistent posteriors
as well. \XPHM{} shows consistency for 22 out of the 30 BBHs. For the remaining eight cases, this model
yields biased posteriors that could be mistaken as violations of GR. 
The more recent phenomenological model \textsc{IMRPhenomXO4a} offers a discernible improvement, but still points
to false GR violations based on the $\Delta a_f/\bar{a}_f$ posteriors for four cases (see Fig.~\ref{fig:IMR_comparison_XPHM_XO4a}).
For the single-spin precessing $Q=8$ BBHs, no single waveform model can be relied on for IMR consistency
though the combined \SEOB-\TPHM{} posteriors mostly yield consistent results.
Additional injections using the new $Q=8$ double-spin precessing simulations of Ref.~\cite{Scheel:2025jct} should be conducted to further test the combined \SEOB-\TPHM{} performance using at least two different cutoffs.
What we can state with confidence is that strong claims of GR violations based on the IMR consistency test
should be based on more than one reasonable frequency cutoff and should employ at least two waveform models, preferably more.
Finally, waveform systematics should always be checked using state of the art NR simulations.

\section{Results III: Incorporating Model Uncertainty into the Injection Study}
\label{Sec:NR_informed_PE}

As introduced in Sec.~\ref{Sec:injection_primer}, the model-dependent posterior distribution
is calculated through Bayes' theorem using Eq.~\eqref{eq:bayes}. If we have and ensemble of
$N$ models and we wish to produce a single distribution that marginalizes over model 
uncertainty, Bayesian model averaging can be used,
\begin{equation} \label{eq:BMA}
	p(\boldsymbol{\lambda} \vert d) =  \sum_{i = 1}^{N} p(\boldsymbol{\lambda} \vert d, \model_{i}) p(\model_{i} | d),
\end{equation}
where $p(\model_{i} | d),$ is the probability of the model $\model_{i}$ given the data, and
$p(\boldsymbol{\lambda} \vert d, \model_{i})$ is the model-dependent posterior distribution
introduced in Eq.~\eqref{eq:bayes}. By re-applying Bayes' theorem to Eq.~\eqref{eq:BMA}, it can
be shown that Bayesian model averaging simply averages the model-dependent posterior 
distributions, weighted by their respective evidence~\cite{Ashton:2019leq}, hereafter referred
to as the \textit{evidence informed approach}:
\begin{align} \label{eq:evidence_informed}
	p(\boldsymbol{\lambda} \vert d) &=  \sum_{i = 1}^{N} \left[ \frac{\mathcal{Z}_{i} \Pi(\model_{i})}{\sum_{j=1}^{N} \mathcal{Z}_{j} \Pi(\model_{j})}\right]  p(\boldsymbol{\lambda} \vert d, \model_{i}), \\ \nonumber
	&=  \sum_{i = 1}^{N} w_{i} p(\boldsymbol{\lambda} \vert d, \model_{i}), 
\end{align}
where $\Pi(\model_{i})$ is the discrete prior probability for the choice of model, which is often assumed to constant and equal to $1 / N$, and $w_{i}$ is the weight given the each model dependent posterior distribution. Alternatively, we can remain completely agnostic and simply mix the model dependent posterior distributions with equal weight (see e.g. Ref.~\cite{LIGOScientific:2021djp}), i.e. $p(\model_{i} | d) = 1 / N$, hereafter referred to as the
\textit{standard approach}.

An alternative technique for marginalizing over model uncertainty is to simultaneously infer
the model and model properties in a single \textit{joint} Bayesian
analysis~\cite{Hoy:2022tst}. This technique expands the binary properties
$\boldsymbol{\lambda} = \{\lambda_{1}, \lambda_{2}, ..., \lambda_{N}\}$ to include an
additional parameter, $m$, which is mapped to a model $\model_{i}$ during
the sampling. As with other parameters, a prior must be defined for choice of model.

Defining a prior probability for the choice of model is challenging as the accuracy of each
model varies across the parameter space. For this reason Ref.~\cite{Hoy:2024vpc} proposed
using a parameter-space dependent prior informed by the model's accuracy to numerical
relativity simulations. This implies that the most accurate GW model will more likely be used
to evaluate the likelihood in each region of parameter space. Although other functional
forms are possible, Ref.~\cite{Hoy:2024vpc} suggested the following prior,
\begin{equation}
	\Pi(\model_{i} | \boldsymbol{\lambda}) \propto \mismatch_{i}(\boldsymbol{\lambda})^{-4}
\end{equation}
where $\mismatch_{i}$ is the mismatch for model $\model_{i}$ as defined in 
Eq.~\eqref{eq:mismatch}. Unfortunately, calculating the mismatch exactly is computationally 
expensive. Consequently, an interpolant can be built prior to the analysis
which is fast to evaluate~\cite{Hoy:2024vpc}. In this work, we employ this technique and use
the same interpolant as introduced in Ref.~\cite{Hoy:2024vpc}. We hereafter refer to this
algorithm as the \textit{NR-informed method}.

The mismatch interpolant constructed in Ref.~\cite{Hoy:2024vpc} has only been
trained on binaries with mass ratios $1/4 \le q \le 1$. Therefore, in this section we focus only
on simulations whose intrinsic and extrinsic parameters are listed in Tables \ref{tab:SXS_sim_params} 
and \ref{tab:SXS_extrinsic_params}.

%==============================================================
%    FIG: pulse plots: NR vs Evidence-informed vs Standard 30 sims, 4 parameters
%==============================================================
\begin{figure*}[tp!]
    \centering
    \includegraphics[width=0.99\textwidth]{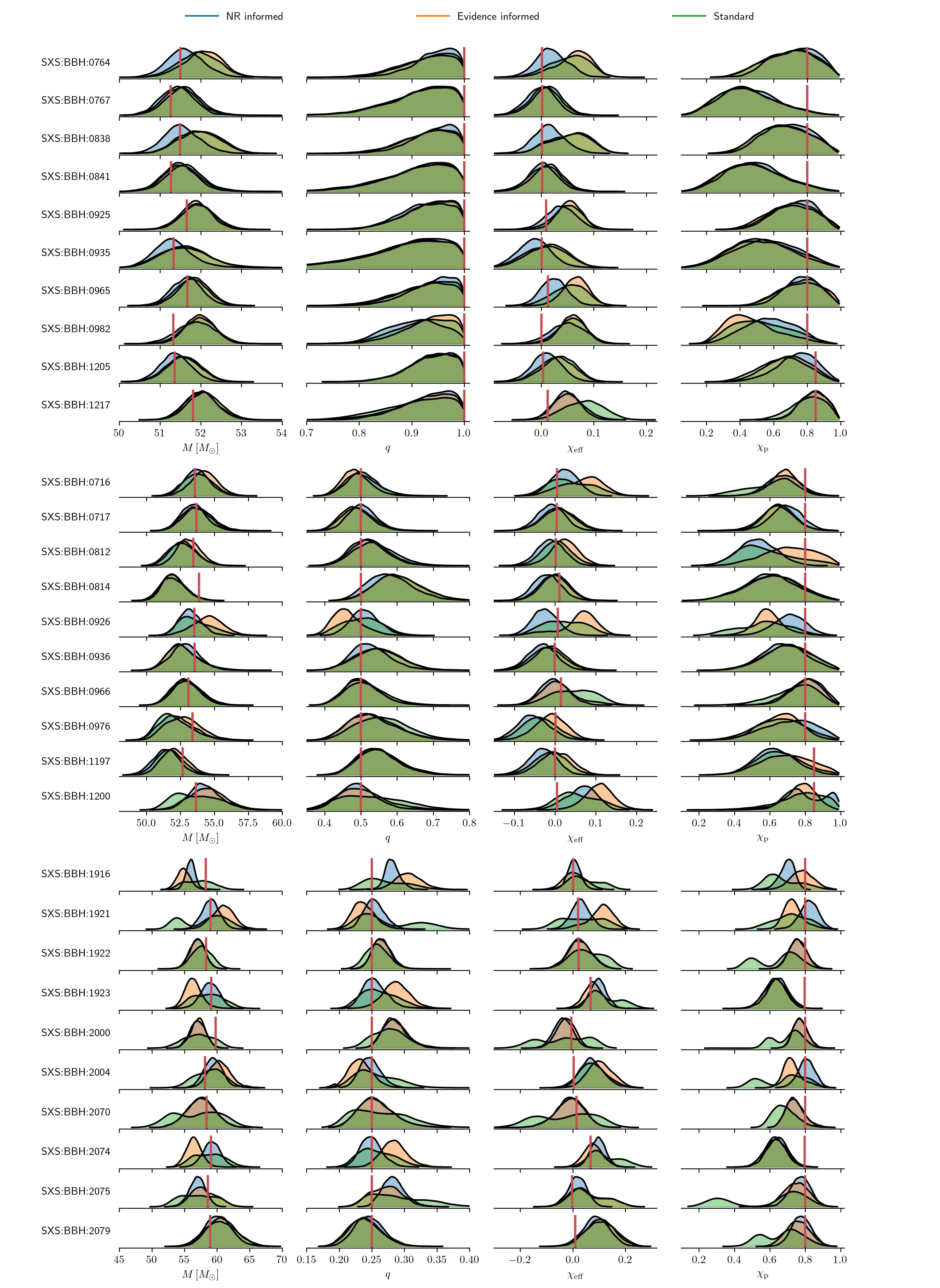}
\caption{One-dimensional marginalized posterior distributions obtained for the inferred total mass (first column), mass ratio (second column), effective parallel (third column) and effective perpendicular spins (fourth column) for the 30 \SXS{} binary black hole simulations used throughout this work. The red vertical lines indicate the true values. The $\{$blue, orange, green$\}$ posteriors are obtained via the NR-informed, evidence-informed and standard methods for combining PE results from multiple waveform 
models (see Sec.~\ref{Sec:NR_informed_PE}).}
\label{fig:three_approach_posteriors}
\end{figure*}

\begin{table*}[t!]
\begin{center}
 \begin{tabular}{lccc|ccc|ccc|ccc}
 \hline
\hline
$\Theta$	 & 	$\rat_\text{av}^\textsc{N} $	 & 	$\rat_\text{av}^\textsc{E} $	 & 	$\rat_\text{av}^\textsc{S} $	 & 	$\rat_\text{av}^\textsc{N}[Q=1] $	 & 	$\rat_\text{av}^\textsc{E}[Q=1] $	 & 	$\rat_\text{av}^\textsc{S}[Q=1] $	 & 	$\rat_\text{av}^\textsc{N}[Q=2] $	 & 	$\rat_\text{av}^\textsc{E}[Q=2] $	 & 	$\rat_\text{av}^\textsc{S}[Q=2] $	 & 	$\rat_\text{av}^\textsc{N}[Q=4] $	 & 	$\rat_\text{av}^\textsc{E}[Q=4] $	 & 	$\rat_\text{av}^\textsc{S}[Q=4] $	\\
\hline
$ \Mtot $	 & 	0.827	 & 	0.750	 & 	0.845	 & 	0.922	 & 	0.813	 & 	0.808	 & 	0.810	 & 	0.828	 & 	0.834	 & 	0.749	 & 	0.610	 & 	0.892	\\
$ q $	 & 	0.748	 & 	0.691	 & 	0.769	 & 	0.553	 & 	0.559	 & 	0.556	 & 	0.905	 & 	0.851	 & 	0.867	 & 	0.787	 & 	0.662	 & 	0.885	\\
$ \chi_\text{eff} $	 & 	0.804	 & 	0.763	 & 	0.793	 & 	0.846	 & 	0.672	 & 	0.692	 & 	0.763	 & 	0.832	 & 	0.889	 & 	0.802	 & 	0.785	 & 	0.797	\\
$ \chi_\text{p} $	 & 	0.698	 & 	0.675	 & 	0.633	 & 	0.713	 & 	0.674	 & 	0.667	 & 	0.675	 & 	0.669	 & 	0.650	 & 	0.707	 & 	0.681	 & 	0.583	\\
\hline
\hline
\end{tabular}
\caption{The recovery score \eqref{eq:ratio_std_to_cost} 
of each method for combining posteriors from different waveform models for the parameters listed in column one.
Columns two to four list the averages over the entire set of 30 simulations for NR-informed, evidence-informed and the standard methods respectively labeled as \textsc{N, E, S} here. The remaining columns show the average of the same quantity
over the size-10 $Q=1,2,4$ subsets. A value of $\rat >0.7$ indicates a recovery of the
injected parameter within one standard deviation of the mean of the posterior distribution.
Note that the lower recovery score for the mass ratio $q$ for the $Q=1$ subset is due to the fact that the injected value
lies at the boundary of the domain hence causing the posteriors to rail as can be seen in Fig.~\ref{fig:three_approach_posteriors}. The closer the recovery score to unity, the better.}\label{tab:recovery_scores_three_methods}
\end{center}
\end{table*}

We present the main results in Fig.~\ref{fig:three_approach_posteriors} where
we recover the 30 injections of Table~\ref{tab:SXS_sim_params} using the
three methods summarized above. Note that the color coding has now changed:
blue corresponds to the NR-informed method \cite{Hoy:2024vpc}, green to the evidence-informed method \cite{Ashton:2019leq}
and orange to the standard method \cite{LIGOScientific:2021djp} for combining posteriors from multiple waveform models.
For most of the panels in the figure, we see that
the three approaches mostly yield overlapping posteriors. In what follows, we will focus
on cases (or parameters) where a discernible disagreement is observed.
The figure is further supplemented by the recovery scores given in Table~\ref{tab:recovery_scores_three_methods} presented in the same format as Table~\ref{tab:recovery_scores}.

Starting with the $Q=1$ subset, we see that the NR-informed method produces slightly
better posteriors for the total binary mass $\Mtot$.
The $q$ posteriors strongly overlap with each other with the most prominent disagreement
occurring for \sxsSim{0982} which proved to be a very challenging injection for
all three waveform models as exhibited in Fig.~\ref{fig:three_model_posteriors}.
The NR-informed method shows more discernible improvement compared to the other two methods
for inferring $\chieff$. On the other hand, the $\chip$ posteriors are mostly indistinguishable
except for two cases where the NR-informed method produces slightly better posteriors.
These findings are quantified in Table ~\ref{tab:recovery_scores_three_methods} 
in terms of the recovery score, $\rat$, once again averaged over the 10 $(Q=1)$ simulation posteriors for each parameter of interest.
We also note that the $(Q=1)$-average recovery scores produced by the NR-informed method are also
slightly better (closer to unity) than the best recovery score produced by an individual waveform model as 
given in Table~\ref{tab:recovery_scores} with the exception of $q$ posteriors, 
which are virtually the same between the three different methods.

For the $Q=2$ subset, the standard method seems to yield a slightly better average recovery score for the total binary mass $M$ than the
other two methods. This is partly due to the aforementioned property of the cost function \eqref{eq:cost_function} penalizing asymmetric posteriors more severely, which is most
evident for \sxsSim{1200}. Visually, the NR-informed (blue) posterior seems like a better
recovery than the standard (green) one, but as the former is much less symmetric than the latter, it returns
a higher cost function with respect the corresponding standard deviation.
The evidence-informed method also produces a marginally higher average
recovery score for $M$ than the NR-informed one. However, the latter method does yield the highest
values of $\rat$ for the mass ratio $q$ as can be corroborated by Table~\ref{tab:recovery_scores_three_methods} and Fig.~\ref{fig:three_approach_posteriors}.
In fact, $\rat_\text{av}$ is once again higher than the highest average recovery produced by an individual waveform model (see Table~\ref{tab:recovery_scores}).
Overall, all three methods yield recovery scores above $0.7$ for all $Q=2$ simulations except for one:
\sxsSim{0814}. This is not surprising as none of the waveform models can recover the injected value $q=1/2$ within $\pm \sigma$ for this
case (see Fig.~\ref{fig:three_model_posteriors}).

As for the recovery of the effective spins, the standard method yields the highest
recovery scores for $\chieff$ owing to the fact that it produces the widest and the most symmetric
posteriors for this parameter.
Given that for $Q=2$, it was the model \XPHM{} that yielded the highest recovery scores, we
expect that neither the NR-informed, nor the evidence-informed methods produce the best recoveries
since these approaches tend to mostly prefer \SEOB{} and/or \TPHM{} over \XPHM{} when generating
their respective posteriors.
For the recovery of $\chip$, the three methods produce comparable recovery scores with the NR-informed
method yielding marginally the highest average recovery score of $\rat_\text{av}=0.675$ close to the best average produced by \SEOB{} ($\rat_\text{av}=0.702$). This outcome is consistent
assuming that both the NR-informed and evidence-informed methods heavily favour \SEOB{} as both \TPHM{}
and \XPHM{} yield $\rat_\text{av}\approx 0.3$. The standard method, on the other hand,
produces wider posteriors, thus also capturing the injected values within $1\sigma$ for most cases.

For the $Q=4$ subset, the standard method yields the highest recovery scores owing to the fact
that it yields the widest posteriors for $M$ and $q$, some of which are bimodal due to the fact
that the underlying waveform models produced strongly disagreeing posteriors as can be seen from the bottom left section of Fig.~\ref{fig:three_model_posteriors}. This is most evident for \sxsSim{1921} and \texttt{2070}. The NR-informed method, on the other hand, produces narrower posteriors, which recover the injections well
for eight of the 10 cases except for \sxsSim{1916} and \texttt{2000}.
For the latter, the method's posteriors are most similar to \SEOB's as it employs \SEOB{} 98.4\% of the time in the parameter
estimation analysis due to the fact that \SEOB{} yields the lowest mismatches to NR in the relevant parameter space. And though $\{M,q\}$ recovery is less than ideal, $M_c$ is recovered
within $1\sigma$ and $\eta$ within $2\sigma$. In this sense, the recovery is still robust
since $M_c$ and $\eta$ are the parameters most relevant to the waveform phase.

For the case of \sxsSim{1916}, we see that the NR-informed analysis infers $\{M,q\}$ posteriors that are similar to \SEOB's, but not identical. This is despite the fact that the method uses \SEOB{} 94.7\% of the time (\TPHM{} 4.7\% and \XPHM{} 1\%).
As a first check,  we re-ran this PE analysis with more aggressive settings and obtained statistically indistinguishable posteriors indicating that the run results have converged.
As can be gathered from Figs.~\ref{fig:three_model_posteriors} and \ref{fig:three_approach_posteriors}, the NR-informed analysis performs better than the \SEOB-only analysis when estimating the spins, but  worse when looking at the $M$-$q$ ($m_1$-$m_2$) projection of the 15D parameter space.
However, when we translate the $M$-$q$ posteriors to $M_c$-$\eta$ posteriors, we observe that
the chirp mass is recovered with a score of $\rat > 0.9$ and the difference between the injected value
of $\eta$ and the median of the posterior is $\approx 5\%$.
In short, though one may naively expect the NR-informed method to agree with \SEOB{} for this case, given 
\begin{inparaenum}[(i)]
 \item
 the high dimensionality of the parameter space, 
 \item 
 the stochastic nature of the parameter estimation process
and 
\item 
the fact that the NR-informed method does not use \SEOB{} 100\% of the time,
\end{inparaenum}
 the most likely parameters  need not agree with those from an \SEOB{} PE run.

Another interesting result of note is that the evidence-informed method yields posteriors for \sxsSim{1916} that
are very similar to those from the \XPHM{} run, consistent with the fact that \XPHM{}
has overwhelmingly the largest Bayesian evidence for this simulation: when computing the weights in Eq.~\eqref{eq:evidence_informed}, we find that evidence-informed gives a mixing fraction of 99.99\% for \XPHM.
In a similar manner, this method gives a mixing fraction of 99.99\% for \SEOB{} for \sxsSim{2000}, hence
the similarity in the posteriors between \SEOB{} and the evidence-informed method for this case.
We find it intriguing that this method so heavily favors two different models for two seemingly very similar sets of intrinsic parameters (however, the extrinsic parameters differ significantly).

When it comes to inferring spin information, the standard method outputs the
most uninformative posteriors exhibiting bimodalities and even trimodalities in the recovery of $\chieff$.
Therefore, its slightly better recovery score compared to the evidence-informed method should be
taken in this context. The method's recovery of $\chip$ further degrades due to the fact
that the individual model posteriors overlap weakly resulting in a 90\% CI width of $\approx 0.5$
for the standard method for injected values of $\chip\simeq 0.8$.
On the other hand, both the NR and evidence-informed methods recover the injections well with
the former yielding $\rat_\text{av}\simeq 0.8, 0.7$, respectively for $\chieff, \chip$, and the latter
method giving slightly lower values. The average recovery scores coming from the NR-informed method
further equal or marginally exceed the best $Q=4$ recovery scores from Table~\ref{tab:recovery_scores}.

In short, our findings indicate that for mass ratios of $1:1$ and $4:1$, the NR-informed method
provides the most consistent parameter recovery with respect to the injections.
For mass ratios of $2:1$, this method still gives the most reliable recovery for $q$ and $\chip$
with its recovery of the total mass $M$ only slightly inferior to the other two methods.
This is because the method mostly employs the waveform model \textsc{SEOBNRv5PHM} 
as it is the most accurate (non-surrogate) model for most
of the parameter space of precessing BBHs \cite{MacUilliam:2024oif}.
However, it does not recover $\chieff$ as well as \textsc{IMRPhenomXPHM}
for the $Q=2$ set of simulations injected here. As a result, the NR-informed method does not
provide the best recovery for $\chieff$.
This is one of the main takeaways of this section: the most accurate waveform model does not
necessarily yield the most accurate posteriors for a single parameter, but it is likely to provide a more accurate 15-dimensional posterior distribution. 

It is also possible that the accuracy of the NR-informed method is being compromised by the limitations of
the interpolant used to predict model faithfulness to NR in the parameter space.
As detailed in Ref.~\cite{Hoy:2024vpc}, the interpolant is rather coarse: it is constructed from 250 points
in the effective spin space and on average it is 5\% accurate to the logarithm of the true mismatch.
Therefore, it remains a possibility that the method may sometimes favour the less accurate waveform model in certain
parts of the parameter space, although we did not find this to be the case for the BBHs analysed in Ref.~\cite{Hoy:2024vpc}.

\section{Summary and Outlook}
\label{Sec:discussion}
In this article, we tested the robustness of three state of the art waveform models, \textsc{SEOBNRv5PHM, IMRPhenomTPHM} and \textsc{IMRPhenomXPHM}, by performing Bayesian inference on 35 strongly precessing numerical relativity simulations. 30 of these simulations featured BBHs with significant spin magnitudes: $\{a_1,a_2\} \in [0.8,0.85]$,
nearly planar spin tilts: $\vert \theta_i-\pi/2 \vert \le 10^\circ, i=1,2$, and mass ratios of $1:1$, $2:1$ and $4:1$. The remaining five simulations were systems with single spin precession and mass ratio of $8:1$. We arranged the extrinsic parameters in each case such that the total SNR in an advanced LIGO O4 network consisting of the Hanford and Livingston observatories equaled 40 with a precession SNR of 10.

We first focused on the recovery performance of the individual waveform models as detailed in Sec.~\ref{Sec:recovery_3models}. 
For the sake of brevity, we presented one-dimensional (marginalized) model posterior distributions for only four intrinsic parameters: the total mass $M$,
the small mass ratio $q$, the effective inspiral spin $\chieff$ and the effective precession spin $\chip$.
The full model dependent posterior distribution can be found in our open-access data repository\footnote{\url{https://github.com/akcays2/Injection_Campaign}}.
We then looked at model performance under the inspiral-merger-ringdown consistency test in Sec.~\ref{Sec:IMR_tests}, specifically documenting this in terms of the posteriors for $\Delta M_f/\bar{M_f}$
and $\Delta a_f/\bar{a}_f$ given by Eqs.~\eqref{eq:M_f_bar} - \eqref{eq:Delta_a_f}.
We additionally investigated how the results of this test depend on the choice of the
cutoff frequency.
Finally, we compared three different methods for combining parameter posteriors from multiple
waveform models in Sec.~\ref{Sec:NR_informed_PE}.
We presented the results from this comparison once again in terms of the marginalized 1D posteriors of $\{M, q, \chieff,\chip\}$.

Though we noted many interesting findings from our study throughout this article, we summarize  
our most robust conclusions and recommendations below. However, we note that as with any injection-recovery analysis our conclusions are dependent on the simulations analysed, i.e., our specific choice of strongly precessing systems, on the values of extrinsic parameters selected and on the chosen waveform models.
\begin{itemize}
\item
\SEOB{} provides the least biased parameter recovery for systems with mass ratios less than or equal to $4:1$. 
Though we reported a few occurrences of biases, its overall mean
recovery scores for $\{M, q, \chieff,\chip\}$ testify that, on average, it captures each injection within one standard deviation of the posterior median.
\item
\TPHM{} and \XPHM{} perform reasonably well for the systems with mass ratios of $1:1$ and $2:1$. As indicated by their recovery scores given in Table~\ref{tab:recovery_scores}, we see that the true source properties are mostly captured within one standard deviation of the median.
However, these waveform models produce a biased inference for $\chi_{\mathrm{p}}$ for mass ratios of $4:1$.
We understand that this is because these models infer biased estimates for the spin magnitudes of each black hole, although they recover the spin tilt angles well.
This is also the case for \SEOB{} for the few systems where it exhibits biases in $\chip$.
\item
\new{One likely cause of biases in the $\chip$ inference is the multipole symmetry \eqref{eq:h_coprec_symmetry} that the models employ. This is only an approximation
in the case of co-precessing multipoles and needs to be abandoned to improve the models.
}
 \item
The waveform models yield much more biased posteriors for BBHs with mass ratios of $8:1$.
However, the combined \SEOB-\TPHM{} posteriors capture $\{M,q,\chieff\}$ well.
Nonetheless, in light of the new catalog of numerical relativity simulations, we recommend that this should be
further tested with other double-spin precessing simulations.
\item
The results of the IMR consistency test applied to waveforms with multiple harmonics
emitted by strongly precessing systems, depend on the choice of the cutoff frequency.
\item
The use of the Schwarzschild ISCO frequency as the cutoff [Eq.~\eqref{eq:Sch_ISCO_freq}, Table~\ref{tab:GR_quantiles}] yields
narrower posteriors for $\Delta M_f/\bar{M_f}$ and $\Delta a_f/\bar{a}_f$, with
some cases displaying false GR violations using the waveform model \XPHM{}  (\sxsSim{0838, 0965, 1922, 2000, 2070} according
to the criteria introduced in Sec.~\ref{sec:IMR_test_Qle4}).
On the other hand, using the Kerr ISCO frequency for the cutoff [Eq.~\eqref{eq:Kerr_ISCO_freq}, Table~\ref{tab:GR_quantiles}] leads to at most two GR violations (\sxsSim{1921, 0716}), albeit with broader posteriors for $\Delta M_f/\bar{M_f}$ and $\Delta a_f/\bar{a}_f$.
 \item 
 Regardless of the cutoff, for systems with mass ratios ranging up to $4:1$,
 \TPHM{} is the most reliable model for passing the IMR consistency test with \SEOB{} coming a close
 second. 
  \item
 Model performance under the IMR consistency test degrades as we go from mass ratio $4:1$ systems
 to $8:1$ systems.
 This is especially evident for the results based on the Schwarzschild ISCO frequency cutoff.
 \item
 The recent phenomenological waveform model \textsc{IMRPhenomXO4a}
 performs better than \XPHM{} in the IMR consistency test as shown in Fig.~\ref{fig:IMR_comparison_XPHM_XO4a}, but not as well as \SEOB{} or \TPHM.
 We also found instances where \textsc{IMRPhenomXO4a} produced more biased 1D posteriors compared to \XPHM{} as exhibited in Fig.~\ref{fig:posterior_comparison_XO4a_with_XPHM} in App.~\ref{Sec:App_A}.
 \item
 Based on the results of our IMR consistency tests, we recommend that at least two waveform models be employed to mitigate against potential issues caused by waveform systematics.
 We also advocate the use of at least two frequency cutoffs.
 \item
 The NR-informed method of combining posteriors from multiple waveform models (by incorporating model
 accuracy \cite{Hoy:2024vpc}) produces intrinsic parameter posteriors most consistent with the injections for 
 highly precessing systems with mass ratios from $1:1$ to $4:1$.
 \item
 The standard method of combining multi-model posteriors via equal weights leads to wide, and often multi-modal distributions when individual waveform models disagree. This is most evident when inferring the precession spin $\chip$ for mass ratio $4:1$ systems.
\end{itemize}

We also highlighted some specific simulations that may be useful for testing the next generation of waveform models:
we suggest performing Bayesian inference on the \sxsSim{1923} and \texttt{2074} simulations.
For IMR consistency tests, we recommend analysing the eight simulations first highlighted in Fig.~\ref{fig:IMR_comparison}.
For testing future methods of combining multiple model-dependent posterior distributions, or techniques that marginalize over waveform uncertainty~\cite[see e.g.][]{Read:2023hkv,Pompili:2024yec,Khan:2024whs,Bachhar:2024olc,Mezzasoma:2025moh,Kumar:2025nwb}, we recommend analysing the injections \sxsSim{0764, 0926, 1200, 1916, 1923, 2004}.
Given that the \SXS{} catalog of numerical relativity simulations has recently been updated and expanded \cite{Scheel:2025jct},
the recommended list of simulations will inevitably change in the future.

\new{Another obvious extension to our work here is to systematically document the
biases caused by the co-precessing multipole symmetry \eqref{eq:h_coprec_symmetry}.
Though we looked at this using the model \textsc{IMRPhenomXO4a} for a few cases, a more
thorough way is to employ the asymmetrized version of \textsc{SEOBNRv5PHM} \`{a} la
Ref.~\cite{Estelles:2025zah} for all our injections.}

Our demonstration of the dependence of the IMR consistency
test on the frequency cutoff brings up the question of whether a single universally meaningful cutoff exists.
Though the ISCO frequency of a spinning (Kerr) black hole seems like a more appropriate
choice than the Schwarzschild value, the former can sometimes be larger than the QNM
frequencies of the final black hole. An alternative choice for the cutoff frequency
could be obtained using the inverse adiabaticity parameter,
$Q_\Omega:=\Omega^2/\dot{\Omega}$ where $\Omega$ is the orbital frequency\footnote{This quantity is related to the more ubiquitous $Q_\omega$, first introduced in Ref.~\cite{Baiotti:2010xh}, via $Q_\Omega\approx Q_\omega/2=\omega^2/(2\dot{\omega})$, where $\omega=2\pi f$.}.
When $Q_\Omega\gg 1$, the binary is in the inspiral regime. When $Q_\Omega \approx \ord(1)$, the transition to plunge takes place.
Our suggestion is to use $f(t_\mathcal{C})$ as the new cutoff where $t_\mathcal{C}$
is the root of $Q_\Omega(t)=\mathcal{C}$.
Somewhat arbitrarily picking $\mathcal{C}=10$ yields GW frequencies between
$80\,$Hz and $100\,$Hz for the 30 \SXS{} simulations of Table~\ref{tab:SXS_sim_params}.
These values are slightly higher than those given by the Schwarzschild ISCO frequency
in Table~\ref{tab:GR_quantiles} so they may offer a reasonable alternative for the cutoff frequency. Of course, a detailed study into a suitable value of $C$ is required.

Although we only focused on three state of the art waveform models that characterize gravitational waveforms from quasicircular binary black holes in this study, gravitational-wave astronomy has plethora of models available, which include different underlying physics, approaches and assumptions. This includes, but not limited to, \textsc{NRSur7dq4}~\cite{Varma:2019csw}, \textsc{TEOBResumS}~\cite{Damour:2014yha, Bernuzzi:2014owa, Nagar:2017jdw,Nagar:2018zoe, Akcay:2018yyh, Akcay:2020qrj, Nagar:2020pcj, Gamba:2021ydi, Albertini:2021tbt, Nagar:2022fep, Gonzalez:2022prs, Nagar:2023zxh},
\textsc{TEOBResumS-Dali}~\cite{Chiaramello:2020ehz, Nagar:2021gss, Nagar:2021xnh, Andrade:2023trh, Nagar:2024dzj, Albanesi:2025txj},
{\textsc{SEOBNRv5EHM}}~\cite{Gamboa:2024hli}, {\textsc{IMRPhenomXODE}}~\cite{Yu:2023lml} and {\textsc{IMRPhenomTEHM}}~\cite{Planas:2025feq}. We therefore encourage that a similar injection-recovery campaign is repeated to assess the performance of some (or all of) these additional (and future) models, and understand the level of bias that can be expected for astrophysically interesting binary systems.

\begin{acknowledgments}
We thank Lorenzo Pompili for comments during the LIGO-Virgo-KAGRA internal review
\new{and the anonymous referee for the peer review.}
We especially thank Lorenzo Pompili for the suggestion of re-performing the IMR consistency test for simulations produced with the same model as for the recovery, and Nathan Johnson-McDaniel for the suggestion of increasing the cutoff frequency to the Kerr ISCO radius. We also thank Hector Estelles and Eleanor Hamilton for guidance in calculating the remnant properties with the
{\textsc{IMRPhenomTPHM}} and {\textsc{IMRPhenomXO4a}} waveforms respectively. We are also
grateful to Eleanor Hamilton for their expertise with rotating and plotting numerical relativity waveforms.
We finally thank Laura Nuttall and Jonathan E. Thompson for discussions throughout this project.
SA and JMU acknowledge support from the University College Dublin Ad Astra Fellowship, and CH thanks the UKRI 
Future Leaders Fellowship for support through the grant MR/T01881X/1.  This work used
the Sciama High Performance Compute (HPC) cluster, which is supported by the ICG, SEPNet and the University of Portsmouth.
{\bf Data availability:} \url{https://github.com/akcays2/Injection_Campaign}.
\end{acknowledgments}

\vspace{1em}
\appendix

\section{Additional Tables and Figures}
\label{Sec:App_A}

\begin{table}[t!]
  \begin{center}
             \begin{tabular}{c c c c c c c c c}
            \hline
            \hline
              \texttt{SXS:BBH} & $\rho_{\mathrm{p}}$ & $d_{\mathrm{L}}(\mathrm{Mpc})$ & $\alpha$   & $\delta$   & $\psi$ & $\vartheta_{\mathrm{LN},0}$ & $\vartheta_{\mathrm{JN},0}$ & $\phi_{JL}$  \\
              \hline              
              \texttt{0764} & 10.00  & 301.25 & 0.35 & 0.52 & 0.63 & 1.57 & 1.31 &4.78 \\
\texttt{0767} & 9.98  & 507.56 & 2.44 & 0.52 & 0.94 & 1.05 & 1.07 &2.82                \\
\texttt{0838} & 10.02  & 440.72 & 0.70 & -0.52 & 0.31 & 1.57 & 1.32 &4.77              \\
\texttt{0841} & 10.00  & 314.61 & 0.35 & 1.22 & 0.31 & 1.05 & 1.08 &2.80               \\
\texttt{0925} & 10.01  & 347.30 & 2.09 & 0.52 & 0.63 & 1.57 & 1.80 &2.10               \\
\texttt{0935} & 9.99  & 177.02 & 1.40 & 0.17 & 0.63 & 1.05 & 0.95 &4.94               \\
\texttt{0965} & 10.00  & 73.10 & 2.44 & -0.87 & 0.94 & 1.57 & 1.79 &2.35               \\
\texttt{0982} & 10.04  & 379.71 & 1.05 & -0.87 & 0.94 & 1.57 & 1.39 &4.50              \\
\texttt{1205} & 9.99  & 212.96 & 1.40 & -0.17 & 0.94 & 1.57 & 1.65 &2.74               \\
\texttt{1217} & 10.00  & 378.72 & 1.05 & -0.52 & 0.31 & 1.57 & 1.33 &5.56              \\
\hline              
\texttt{0716} & 10.01  & 374.75 & 0.00 & 0.17 & 0.00 & 1.05 & 1.23 &2.62               \\
\texttt{0717} & 10.00  & 403.21 & 2.09 & 0.87 & 0.00 & 1.05 & 1.33 &1.07               \\
\texttt{0812} & 10.00  & 617.22 & 1.05 & -1.22 & 0.94 & 0.52 & 0.27 &5.21              \\
\texttt{0814} & 10.00  & 855.66 & 2.79 & 0.52 & 0.00 & 0.20 & 0.58 &1.34               \\
\texttt{0926} & 9.99  & 440.74 & 0.00 & -1.22 & 0.00 & 1.57 & 1.94 &1.54               \\
\texttt{0936} & 9.99  & 374.01 & 1.40 & -0.17 & 0.63 & 0.52 & 0.32 &5.74               \\
\texttt{0966} & 10.00  & 133.31 & 2.79 & -0.87 & 0.63 & 1.05 & 0.74 &3.90              \\
\texttt{0976} & 10.00  & 583.42 & 0.70 & 1.57 & 0.31 & 0.52 & 0.78 &1.18               \\
\texttt{1197} & 10.00  & 311.98 & 1.05 & 0.17 & 0.63 & 0.52 & 0.85 &1.77               \\
\texttt{1200} & 10.00  & 759.11 & 0.35 & -0.17 & 0.31 & 0.20 & 0.55 &1.94              \\
\hline
\texttt{1916} & 10.00  & 524.54 & 2.44 & -1.57 & 0.00 & 0.20 & 0.80 &1.76              \\
\texttt{1921} & 11.46  & 693.67 & 0.00 & -0.87 & 0.63 & 0.52 & 1.06 &1.17              \\
\texttt{1922} & 9.95  & 168.36 & 2.09 & -0.87 & 0.63 & 1.57 & 1.08 &5.46               \\
\texttt{1923} & 10.12  & 744.55 & 0.35 & -0.52 & 0.94 & 0.20 & 0.53 &1.85              \\
\texttt{2000} & 9.98  & 283.26 & 1.05 & 1.57 & 1.26 & 1.57 & 0.97 &5.21                \\
\texttt{2004} & 10.00  & 263.35 & 1.05 & 0.87 & 0.31 & 1.05 & 0.56 &5.67               \\
\texttt{2070} & 10.02  & 290.78 & 1.75 & 0.87 & 0.63 & 1.57 & 1.69 &2.97               \\
\texttt{2074} & 10.09  & 754.28 & 2.79 & 0.87 & 0.00 & 0.20 & 0.46 &1.69               \\
\texttt{2075} & 10.01  & 468.36 & 3.14 & 0.87 & 0.31 & 1.05 & 1.56 &0.89               \\
\texttt{2079} & 10.27  & 555.25 & 2.09 & 0.87 & 0.31 & 0.20 & 0.75 &1.83               \\              
\hline
              \hline 
          \end{tabular}
     \caption{Extrinsic parameters for the 30 \SXS{} BBH systems for which we provide the relevant intrinsic parameters in Table~\ref{tab:SXS_sim_params}. See Sec.~\ref{Sec:notation} for the definition of each symbol. All angles are given in radians.}\label{tab:SXS_extrinsic_params}
  \end{center}
\end{table}

\begin{table}[t!]
\begin{center}
 \begin{tabular}{ccccc}
 \hline
\hline
  \texttt{SXS:BBH} &  $f_\text{ISCO}^\text{Sch} \text{(Hz)}$ & $f_\text{ISCO}^\text{Kerr}\text{(Hz)}$ & $\mathcal{D}^{\text{2D}}_{\text{Sch}}$ (\%) & $\mathcal{D}^{\text{2D}}_{\text{Kerr}}$ (\%) \\
  \hline
\texttt{0764} & 85 & 200 & 94 & 70 \\
\texttt{0767} & 86 & 190 & 11 & 21 \\
\texttt{0838} & 85 & 199 & 92 & 76 \\
\texttt{0841} & 86 & 191 & 13 & 20 \\
\texttt{0925} & 85 & 205 & 31 & 7 \\
\texttt{0935} & 86 & 190 & 6 & 9 \\
\texttt{0965} & 85 & 205 & 85 & 30 \\
\texttt{0982} & 86 & 196 & 56 & 11 \\
\texttt{1205} & 86 & 194 & 1 & 12 \\
\texttt{1217} & 85 & 217 & 64 & 38 \\
\hline
\texttt{0716} & 82 & 187 & 53 & 89 \\
\texttt{0717} & 82 & 177 & 27 & 59 \\
\texttt{0812} & 82 & 179 & 8 & 44 \\
\texttt{0814} & 82 & 186 & 13 & 40 \\
\texttt{0926} & 82 & 183 & 66 & 39 \\
\texttt{0936} & 82 & 187 & 15 & 21 \\
\texttt{0966} & 83 & 194 & 43 & 23 \\
\texttt{0976} & 82 & 180 & 25 & 63 \\
\texttt{1197} & 84 & 192 & 75 & 53 \\
\texttt{1200} & 82 & 191 & 21 & 52 \\
\hline
\texttt{1916} & 75 & 166 & 73 & 3 \\
\texttt{1921} & 75 & 156 & 39 & 87 \\
\texttt{1922} & 75 & 151 & 90 & 49 \\
\texttt{1923} & 74 & 162 & 30 & 2 \\
\texttt{2000} & 74 & 150 & 84 & 76 \\
\texttt{2004} & 76 & 157 & 23 & 60 \\
\texttt{2070} & 75 & 149 & 87 & 24 \\
\texttt{2074} & 74 & 161 & 65 & 0 \\
\texttt{2075} & 75 & 155 & 18 & 22 \\
\texttt{2079} & 75 & 139 & 65 & 6 \\
\hline
\hline
  \end{tabular}
  \caption{Results from the IMR consistency test for two different choices of cutoff frequency, one based on the Schwarzschild (Sch), $f_\text{ISCO}^\text{Sch}$ [see Eq.~\eqref{eq:Sch_ISCO_freq}], and the other based on the Kerr ISCO frequency, $f_\text{Kerr}^\text{ISCO}$ [see Eq.~\eqref{eq:Kerr_ISCO_freq}]. The probabilities $\mathcal{D}^{\text{2D}}_{\text{Sch}}$ and $\mathcal{D}^{\text{2D}}_{\text{Kerr}}$ denote the fraction of the 2D $p(\Delta M_f/\bar{M}_f, \Delta a_f/\bar{a}_f )$ posterior probability distribution enclosed by the isoprobability contour that passes through the GR value when using the Schwarzschild ISCO and Kerr ISCO as the cutoff frequencies respectively. In this table, smaller probabilities indicate better consistency with GR.}
  \label{tab:GR_quantiles}
\end{center}
\end{table}

\begin{figure}[t!]
    \centering
    \includegraphics[width=0.49\textwidth]{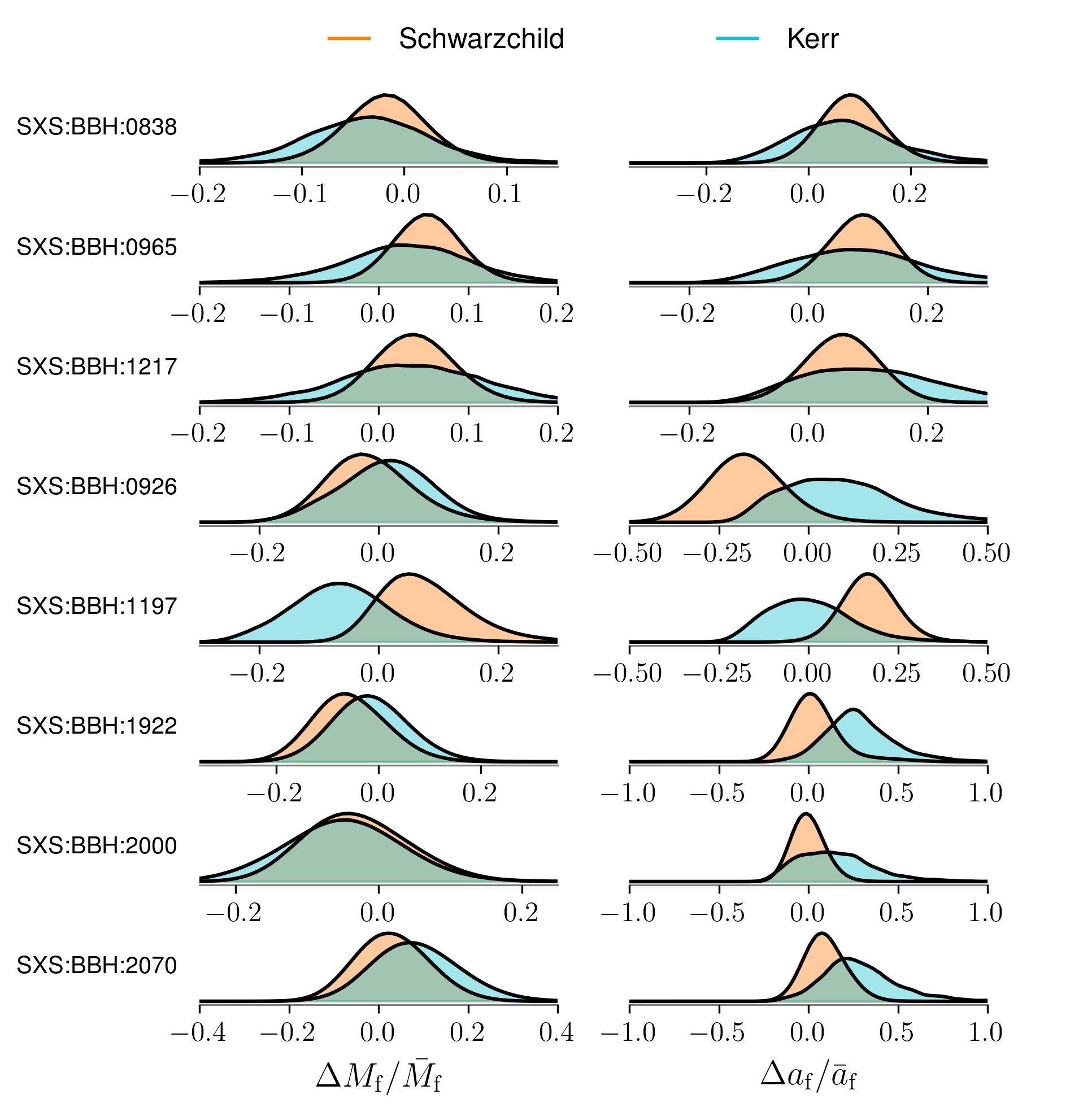}
\caption{\SEOB{}'s performance under the IMR consistency test for the eight special cases
mentioned in Sec.~\ref{sec:IMR_test_Qle4}.
Orange histograms are obtained using the Schwarzschild ISCO as the frequency cutoff
[Eq.~\eqref{eq:Sch_ISCO_freq}] and the cyan ones using the Kerr ISCO frequency given
in Eq.~\eqref{eq:Kerr_ISCO_freq}.
}
\label{fig:IMR_comparison_SEOB}
\end{figure}

\begin{figure}[t!]
    \centering
    \includegraphics[width=0.49\textwidth]{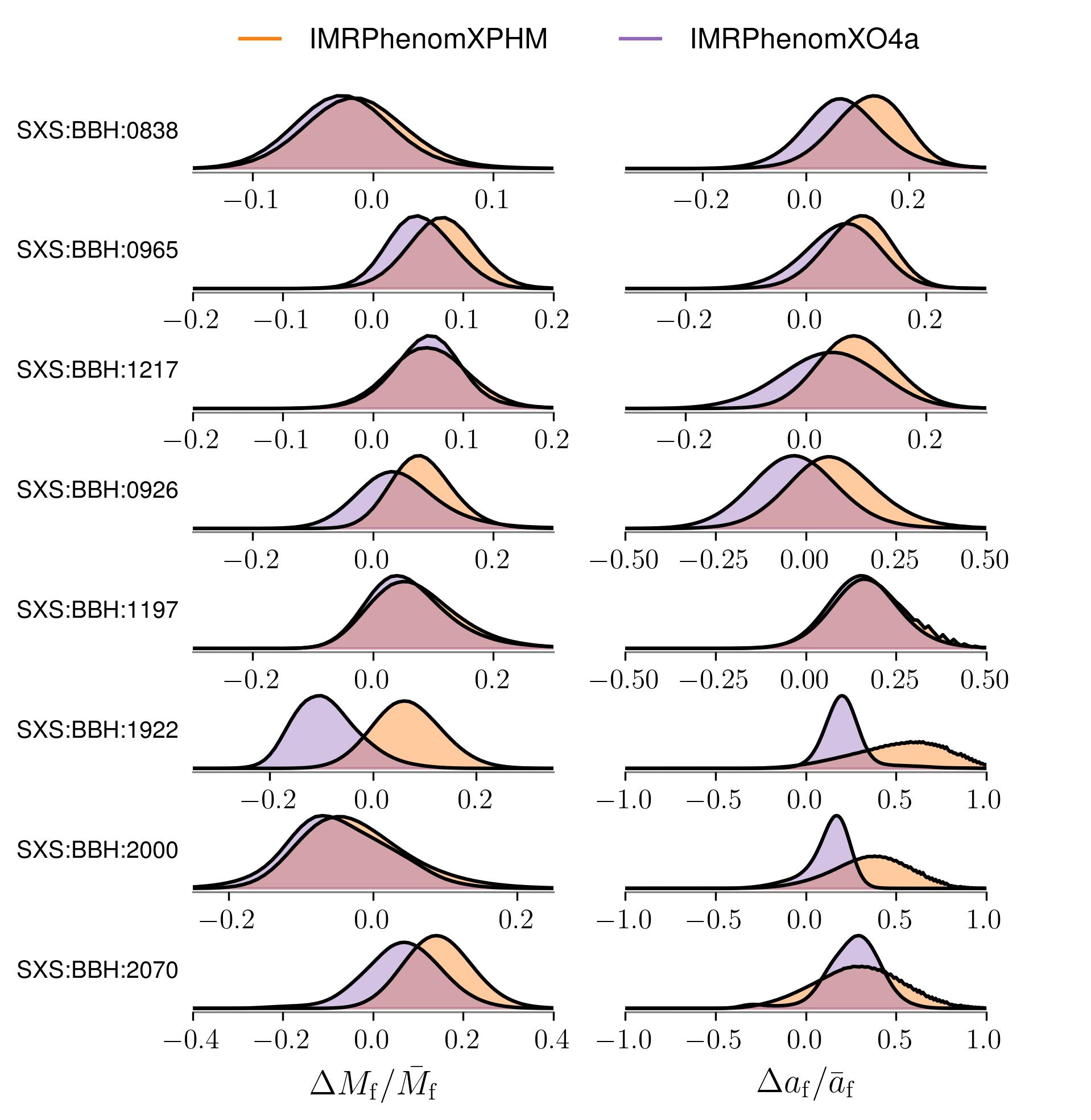}
\caption{
Performance of the \XPHM{} and \textsc{IMRPhenomXO4a} (\XOFourA) models under the IMR consistency test
for eight cases in which \XPHM{} does not recover the true value within two
standard deviations. See Sec.~\ref{sec:IMR_test_Qle4} for details. For these cases we consider a cutoff frequency based on the Schwarzschild ISCO.}
\label{fig:IMR_comparison_XPHM_XO4a}
\end{figure}

\begin{figure}[t!]
    \centering
    \includegraphics[width=0.49\textwidth]{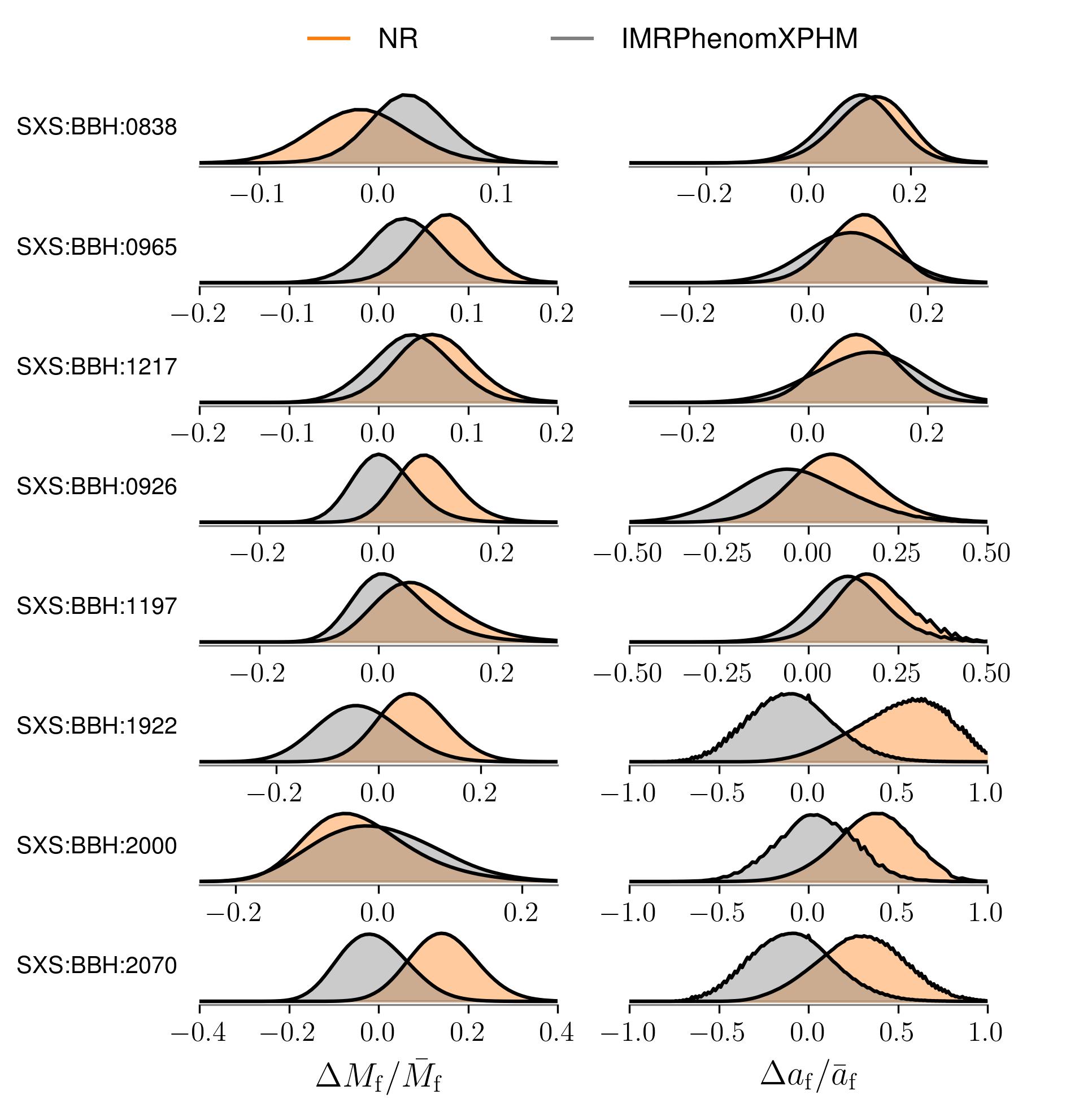}
\caption{Same as Fig.~\ref{fig:IMR_comparison_XPHM_XO4a}, but we show results of the IMR consistency test for NR injections (orange) and \XPHM{} injections with the same parameters as the NR (grey). For both cases, we analyse the injections with \XPHM{} and consider a cutoff frequency based on the Schwarzschild ISCO.}
\label{fig:imr_comparison_nr_vs_xphm}
\end{figure}

\begin{figure*}[t!]
    \centering
    \includegraphics[width=0.98\textwidth]{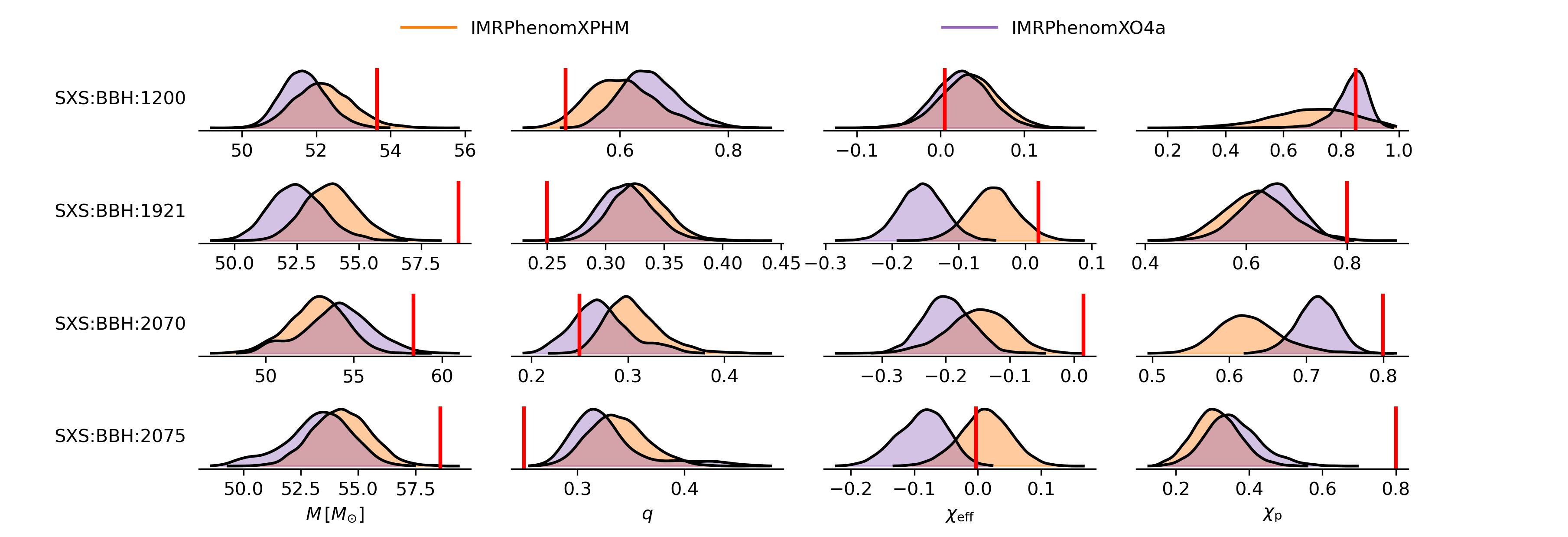}
\caption{Same as Fig.~\ref{fig:three_model_posteriors}, but we specifically compare the performance of the waveform models \XPHM{} and \XOFourA{} for a selection of four \SXS{} binary black hole simulations with mass ratios $Q=2$ or $Q=4$ from Sec.~\ref{sec:Qle4_injections}.
The red vertical lines indicate the true values.}
\label{fig:posterior_comparison_XO4a_with_XPHM}
\end{figure*}

\bibliography{refs, local_bib}

\end{document}